\documentclass[11pt,aps,tightenlines,nofootinbib,longbibliography,superscriptaddress]{revtex4-1}

\usepackage[utf8]{inputenc}
\usepackage{charter} 
\usepackage{fullpage}
\usepackage{hyperref}
\usepackage{graphicx}
\usepackage{lipsum}
\usepackage{hyperref}
\hypersetup{hidelinks}
\usepackage[total={6.5in,9in},top=1in,headsep=0.1in,headheight=1in]{geometry}
\usepackage{enumitem,amssymb}
\usepackage[usenames,dvipsnames]{xcolor}
\usepackage{aas_macros}
\usepackage{xspace}
\usepackage{lineno}

\newcommand\snowmass{
\begin{center}
  \rule[-0.2in]{\hsize}{0.01in}\\
  \rule{\hsize}{0.01in}\\
  \vskip 0.1in
  Submitted to the Proceedings of the US Community Study\\ 
  on the Future of Particle Physics (Snowmass 2021)\\
  \rule{\hsize}{0.01in}\\
  \rule[+0.2in]{\hsize}{0.01in}\\[-2em]
\end{center}
}

\usepackage[firstpage=true]{background}
\backgroundsetup{contents={\parbox{6.5in}{\snowmass}}, scale=1,placement=top,opacity=1,color=black,position={3.25in,1.2in}}

\usepackage{fancyhdr}
\fancypagestyle{plain}{%
  \fancyhf{}%
  \fancyhead[C]{}
  \fancyfoot[C]{\thepage}
}

\fancypagestyle{empty}{%
  \fancyhf{}%
  \fancyhead[C]{{\it Snowmass2021: Observational Facilities to Study Dark Matter}}
  \fancyfoot[C]{\thepage}
}
\begin{document}

\title{Snowmass2021 Cosmic Frontier White Paper:\\ Observational Facilities to Study Dark Matter}

\author{Sukanya Chakrabarti}
\affiliation{School of Physics and Astronomy, Rochester Institute of Technology, 84 Lomb Memorial Drive, Rochester, NY 14623}

\author{Alex Drlica-Wagner}
\affiliation{Fermi National Accelerator Laboratory, P. O. Box 500, Batavia, IL 60510, USA}
\affiliation{Kavli Institute for Cosmological Physics, University of Chicago, Chicago, IL 60637, USA}
\affiliation{Department of Astronomy and Astrophysics, University of Chicago, Chicago, IL 60637, USA}

\author{Ting S. Li}
\affiliation{Department of Astronomy and Astrophysics, University of Toronto, 50 St. George Street, Toronto ON, M5S 3H4, Canada}

\author{Neelima Sehgal}
\affiliation{Physics and Astronomy Department, Stony Brook University, Stony Brook, NY 11794, USA}

\author{Joshua D. Simon}
\affiliation{Observatories of the Carnegie Institution for Science, Pasadena, USA, 91101}

\author{Simon Birrer}
\affiliation{Kavli Institute for Particle Astrophysics and Cosmology and Department of Physics, Stanford University, Stanford, CA 94305, USA}
\affiliation{SLAC National Accelerator Laboratory, Menlo Park, CA, 94025, USA}

\author{Duncan A. Brown}
\affiliation{Department of Physics, Syracuse University, Syracuse, NY 13244, USA}

\author{Rebecca Bernstein}
\affiliation{Observatories of the Carnegie Institution for Science, Pasadena, USA, 91101}

\author{Alberto D. Bolatto}
\affiliation{Department of Astronomy, University of Maryland, College Park, MD  20742, USA}

\author{Philip Chang}
\affiliation{Department of Physics, University of Wisconsin-Milwaukee, 3135 North Maryland Avenue, Milwaukee, WI  53211, USA}

\author{Kyle Dawson}
\affiliation{Department of Physics and Astronomy, University of Utah,
Salt Lake City, UT 84112, USA}

\author{Paul Demorest}
\affiliation{National Radio Astronomy Observatory, P.O. Box O, Socorro, NM  87801, USA}

\author{Daniel Grin}
\affiliation{Department of Physics and Astronomy, Haverford College, 370 Lancaster Avenue, Haverford, Pennsylvania 19041, United States}

\author{David L. Kaplan}
\affiliation{Department of Physics, University of Wisconsin-Milwaukee, Milwaukee, WI  53201, USA}

\author{Joseph Lazio}
\affiliation{Jet Propulsion Laboratory, California Institute of Technology, Pasadena, CA  91106, USA}

\author{Jennifer Marshall}
\affiliation{George P. and Cynthia Woods Mitchell Institute for Fundamental Physics and Astronomy, and Department of Physics and Astronomy, Texas A\&M University, College Station, TX 77843, USA}

\author{Eric J. Murphy}
\affiliation{National Radio Astronomy Observatory, 520 Edgemont Road, Charlottesville, VA  22903, USA}

\author{Scott Ransom}
\affiliation{National Radio Astronomy Observatory, 520 Edgemont Road, Charlottesville, VA  22903, USA}

\author{Brant E. Robertson}
\affiliation{Department of Astronomy and Astrophysics, University of California, Santa Cruz, Santa Cruz, CA, 95060, USA}

\author{Rajeev Singh}
 \affiliation{Institute  of  Nuclear  Physics  Polish  Academy  of  Sciences,  PL-31-342  Krak\'ow,  Poland}

\author{An\v{z}e Slosar}
\affiliation{Physics Department, Brookhaven National Laboratory, Upton NY 11973}

\author{Tommaso Treu}
\affiliation{Department of Physics and Astronomy, University of California, Los Angeles, CA 90095, USA}

\author{Yu-Dai Tsai}
\affiliation{Department of Physics and Astronomy, University of California, Irvine, CA 92697-4575, USA}

\author{Benjamin F. Williams}
\affiliation{Astronomy Department, University of Washington, Seattle, WA  98195, USA}



\begin{abstract}

We present an overview of future observational facilities that will significantly enhance our understanding of the fundamental nature of dark matter. These facilities span a range of observational techniques including optical/near-infrared imaging and spectroscopy, measurements of the cosmic microwave background, pulsar timing, 21-cm observations of neutral hydrogen at high redshift, and the measurement of gravitational waves.
Such facilities are a critical component of a multi-pronged experimental program to uncover the nature of dark matter, while often providing complementary measurements of dark energy, neutrino physics, and inflation.

\end{abstract}

\maketitle


\newpage

\tableofcontents

\newpage

\section{Executive Summary}

The fundamental nature of dark matter represents one of the most compelling open questions in physics.
Since the last Snowmass, the experimental program to determine the nature of dark matter has expanded dramatically \citep[][]{Battaglieri:2017}.
The space of viable dark matter models spans nearly 90 orders of magnitude in particle mass, ranging from ultra-light dark matter ($m \sim 10^{-21}$\,eV) to primordial black holes ($m \sim 10^{68}$\,eV). 
To date, all of the positive, empirical measurements of dark matter (including the aforementioned upper and lower bounds on the dark matter mass) have come from astronomical and cosmological observations. 
Furthermore, the microphysical properties of dark matter (e.g., particle mass, interaction cross section(s), etc.) can be extracted from the macroscopic distribution and clustering of dark matter.
Over the next decade, observational facilities spanning the electromagnetic spectrum, as well as gravitational waves, offer the potential to significantly expand our understanding of dark matter physics.
In this white paper, we discuss the scientific motivation and technological opportunities for observational facilities focused on improving our understanding of dark matter.

This white paper complements other white papers devoted to the description of multi-messenger facilities \citep{Multimessenger-Snowmass}, gamma-ray and X-ray experiments \citep{CF1-GammaExp-Snowmass,CF7-GammaExp-Snowmass}, and gravitational wave facilities \citep{GWFacilities-Snowmass}.

\vspace{1.0em}
\noindent {\bf Key Opportunities}
\begin{enumerate}
    \item  Astrophysical and cosmological observations currently provide the only positive, empirical measurements of dark matter. In the coming decade, observational facilities spanning the electromagnetic spectrum and gravitational waves present an opportunity to significantly advance our understanding of dark matter. 

    \item Proposed millimeter-wavelength facilities, such as CMB-HD, will extend the resolution and sensitivity of cosmic microwave background surveys by factors of five or more. These facilities can determine the number of light thermal species in the universe, constrain interactions between dark matter and standard model particles, and measure the matter power spectrum on small scales using gravitational lensing of the microwave background radiation.
    
    \item Proposed centimeter-wavelength radio observatories, including the ngVLA and DSA-2000, can employ pulsar timing measurements to map the dark matter halo of the Milky Way and the sub-structures it contains.
    
    \item Proposed low-frequency radio experiments, such as LuSEE Night and PUMA, can use the 21-cm line of hydrogen from the Dark Ages through cosmic reionization to probe dark matter physics via the thermal history of intergalactic gas and the timing of the formation of the first stars and galaxies.
    
    \item Proposed optical/near-infrared telescopes, including TMT, GMT, MegaMapper, MSE, and SpecTel, can determine the dark matter halo mass function down to $10^{6}$~M$_{\odot}$ and improve sensitivity to the dark matter particle mass and interaction cross-sections  through observations of gravitational lenses, nearby dwarf galaxies, stars, and stellar streams.
    
    \item Proposed gravitational wave facilities, such as Cosmic Explorer and LIGO-Voyager, can be used to probe dark matter, including axion-like particles in binary neutron star mergers, ultralight bosons through their interactions with black holes or directly as boson stars, dark matter spikes around black holes, and primordial black holes.

    \item Contributions to the technical development and construction of observational dark matter facilities will leverage the core technical and scientific capabilities of the HEP community. Furthermore, the involvement of the HEP community will maximize the scientific output of these facilities. 
    
    \item Many proposed dark matter facilities will also offer complementary capabilities to explore dark energy, neutrino properties, and inflationary physics. The capability to probe dark matter physics should be considered in the design phase of these facilities.
\end{enumerate}

\section{Introduction}

Over the past 50 years, observational facilities funded by the DOE, NSF, NASA, and their international counterparts have produced all of the positive empirical measurements of the fundamental properties of dark matter.  Observational discoveries during this period include the first convincing evidence that dark matter exists, determination of the cosmological abundance of dark matter, measurements of dark matter halos spanning seven orders of magnitude in mass, and measurements of the local dark matter mass distribution within our own Galaxy.  These observations have directly constrained the fundamental microscopic properties of dark matter, such as the minimum dark matter particle mass, the maximum allowed coupling to the Standard Model, and the maximum strength of a wide range of interactions in the dark sector. 
Despite this tremendous progress, the landscape of dark matter models remains wide open, and more powerful observational facilities will be needed to narrow  the allowed parameter space for dark matter, and --- with luck --- conclusively identify the nature of dark matter.

In the last two decades, the high energy physics (HEP) community has been critically important to  advances in astronomical facilities and instruments used to study our universe. 
Development of new instrumentation and technology has been driven by the ambitious goal of extracting fundamental physics from cosmology.
HEP scientists played a crucial role in the construction and operation of the Sloan Digital Sky Survey (SDSS) in the late 1990s and early 2000s \citep{York:2000}. 
The HEP interest in SDSS was driven by the desire to constrain the matter--energy density of the universe, building on the growing connection between quarks and the cosmos \citep{Kolb:1986, Quarks2Cosmos:2003}.
In the subsequent years, HEP observational interests have shifted toward understanding the accelerating expansion of the universe (dark energy).
This interest has motivated unprecedented projects such as the Baryon Oscillation Spectroscopic Survey \cite{boss2013}, the Dark Energy Survey \cite{DES:2005}, the Dark Energy Spectroscopic Instrument \cite{desi2016}, and the Vera C.\ Rubin Observatory \citep{Ivezic:2019}. 
In the coming decade, the Stage~IV ground-based cosmic microwave background experiment CMB-S4 will explore inflationary physics and probe the abundance of light relics in the early universe \cite{Abazajian:2019eic}.
Each of these facilities has made (or will make) crucial contributions to our understanding of dark matter physics.

This white paper focuses on future observational facilities that can be designed and built in the next decade to measure the fundamental properties of dark matter.  For these facilities, we strongly recommend that dark matter physics be included as a fundamental mission goal that guides design decisions and helps sets design requirements.
Although we briefly describe the contributions of current and near-future facilities, we leave a more detailed discussion of those projects to white papers from the scientific collaborations associated with specific facilities.
Furthermore, this white paper is intended to complement other white papers devoted to the description of multi-messenger facilities \citep{Multimessenger-Snowmass}, gamma-ray and X-ray experiments \citep{CF1-GammaExp-Snowmass,CF7-GammaExp-Snowmass}, and gravitational wave facilities \citep{GWFacilities-Snowmass}.

\section{Dark Matter Measurements with Future Facilities}

\begin{figure*}[t!]
\centering
\includegraphics[width=1.0\textwidth]{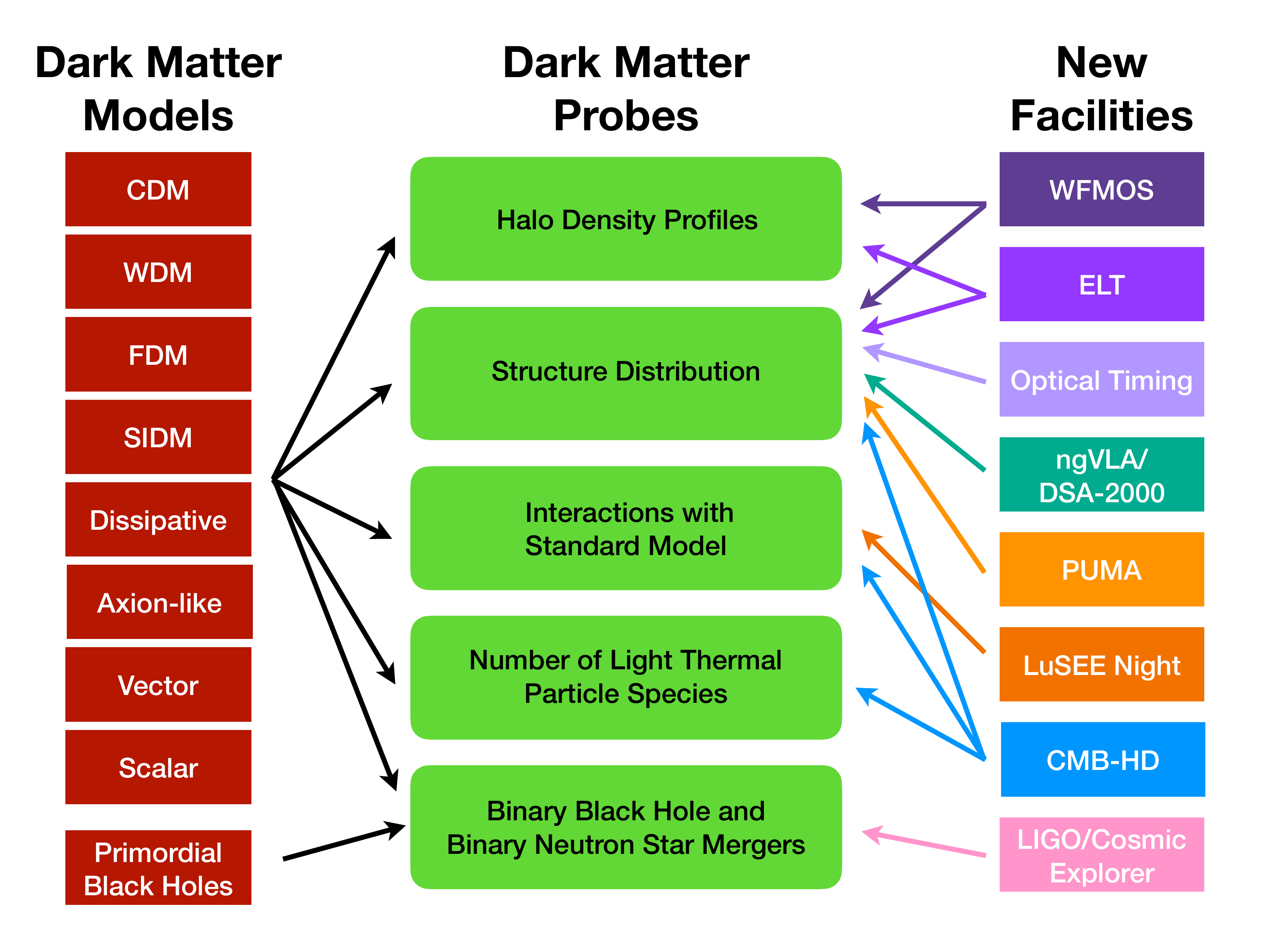}
\caption{Shown are the connections between various dark matter models, the ways that those models affect the observable universe, and the proposed future facilities that will be discussed in this white paper. Acronyms are: CDM: cold dark matter; WDM: warm dark matter; FDM: fuzzy dark matter; SIDM: self-interacting dark matter; WFMOS: wide-field multi-object spectroscopic survey facilities; ELT: extremely large telescope; ngVLA: next-generation Very Large Array; PUMA: Packed Ultrawideband Mapping Array; LuSEE Night: Lunar Surface Electromagnetics Experiment Night; CMB-HD: Cosmic Microwave Background-High Definition; LIGO/Cosmic Explorer: Laser Interferometer Gravitational-Wave Observatory-Voyager and Cosmic Explorer.}
\label{fig:overview}
\end{figure*}

Astrophysical observations have indicated for decades that our Universe contains large amounts of matter that is not made of ordinary baryons.  What this dark matter consists of is one of the biggest unsolved questions in cosmology and particle physics.  Theorists have developed many possible dark matter models, some of which are discussed in ref. \cite{CF7_smallscales}, that could have different astrophysical signatures in regimes that new facilities are aiming to probe. For example, new technology developments are enabling next-generation facilities to probe structure on smaller scales than possible before, and advances in detector technology are enabling unprecedented sensitivities.  Below we outline the key measurement opportunities enabled by next-generation facilities that will probe the properties and nature of dark matter, as depicted in Figure~\ref{fig:overview}.  The proposed facilities that can make these measurements are listed in Table~\ref{table:facilities}.

\begin{table*}
\title{Future Astrophysical Facilities for Dark Matter Science}
\small
\centering
\begin{tabular}{|p{9.5cm}|p{3.2cm}|p{1.7cm}|p{1.5cm}|}
\hline
    Facility & Type & Reference & Section \\
    \hline
    Cosmic Microwave Background-High Definition (CMB-HD) & CMB (mm) & \cite{Sehgal:2019ewc} & \S~\ref{sec:cmb_hd} \\
    next-generation Very Large Array (ngVLA) & radio (mm-cm) &  \cite{Selina2018} & \S~\ref{sec:ngvla_dsa2000} \\
    Deep Synoptic Array-2000 (DSA-2000) & radio (cm) &  \cite{Hallinan2019} & \S~\ref{sec:ngvla_dsa2000} \\
    Lunar Surface Electromagnetics Experiment (LuSEE) Night & 21\,cm (m) &  \cite{Burns2021} & \S~\ref{sec:lusee_night} \\
    Packed Ultrawideband Mapping Array (PUMA) & 21\,cm (cm) &  \cite{Slosar2019} & \S~\ref{sec:puma} \\
    Thirty Meter Telescope (TMT) & optical/near-IR &  \cite{Sanders2013} & \S~\ref{sec_elts} \\
    Giant Magellan Telescope (GMT) & optical/near-IR &  \cite{Johns2012} & \S~\ref{sec_elts} \\
    Dark Energy Spectroscopic Instrument II (DESI-II) & optical &  \cite{Levi2019BAAS} & \S~\ref{sec:desi2} \\
    MegaMapper & optical &  \cite{schlegel19} & \S~\ref{sec:megamapper} \\
    Maunakea Spectroscopic Explorer (MSE) & optical/near-IR &  \cite{MSE_book2018} & \S~\ref{sec:mse} \\
    SpecTel & optical/near-IR &  \cite{SpecTel:2017} & \S~\ref{sec:spectel} \\
    Cosmic Explorer & gravitational wave &  \cite{Evans:2021gyd} & \S~\ref{sec:cosmic_explorer} \\
    LIGO Voyager & gravitational wave &  \cite{LIGO:2020xsf} & \S~\ref{sec:ligo_voyager} \\    
    \hline
\end{tabular}
\caption{Future astrophysical facilities for dark matter science.} \label{table:facilities}
\end{table*}

The challenge currently facing the community is that the landscape of viable dark matter models is vast, with relatively weak guidance provided by theory as to which types of dark matter should be judged as ``best'' or ``most likely to exist.''  In this situation, astronomical facilities offer unique opportunities for constraining many of the particle properties of dark matter:

\begin{itemize}
    \item {\bf Dark matter particle mass.}  The mass ranges for many varieties of dark matter models will be examined by future astronomical facilities.  Optical telescopes can place lower limits on the mass of a warm dark matter particle by measuring the mass function of dark matter halos on subgalactic scales via gravitational lensing or direct acceleration measurements (\S~\ref{sec_elts}) or heating of stellar streams (\S~\ref{sec_wfmos}).  Millimeter-wave CMB observations can be sensitive to warm dark matter and other dark matter models that suppress the matter power spectrum on galactic or sub-galactic scales through gravitational lensing of the CMB; they will also probe axion-like particles over several orders of magnitude of particle mass via multiple mechanisms (\S~\ref{sec_cmb}).  Future pulsar timing facilities, including ngVLA and DSA-2000, will also be able to probe dark matter sub-structure (\S~\ref{sec:pulsartiming}).
    
    \item {\bf Dark matter self-interaction cross-section.} Interactions between dark matter particles, either self-interactions or interactions among multiple species of dark matter, can only be probed with astronomical measurements.  These interactions are most detectable via their impact on the dark matter density at the centers of halos and the internal density structure of halos.  Future spectroscopic survey facilities (\S~\ref{sec_wfmos}) and extreme-precision radial velocity observations enabled by the ELTs (\S~\ref{sec_elts}) are likely to have the largest impact in this area.
    
    \item {\bf Dark matter coupling to standard model particles.}  Coupling between dark matter particles and ordinary matter can produce standard model particles such as photons.  These interactions can be imprinted in the light from the early Universe in both the primordial CMB signal and in tracers of reionization and intergalactic medium heating. These signals can be directly probed by millimeter-wave CMB observations (\S~\ref{sec_cmb}) and global 21\,cm measurements of the dark ages (\S~\ref{sec_21}). In addition, gravitational probes can indirectly constrain dark matter coupling with standard model particles via its impact on structure formation; these tests can be carried out using deep optical observations of the local dwarf galaxy population (\S~\ref{sec_elts}), measurements of the structure distribution on small scales via CMB lensing, optical strong lensing, and direct acceleration measurements (\S~\ref{sec_cmb}, \S~\ref{sec_elts}, \S \ref{sec:pulsartiming}), and the clumping of stellar streams (\S~\ref{sec_wfmos}).
    
    \item {\bf Dark sector properties.} Many dark-sector models contain new light particles that were once in thermal equilibrium with standard model particles.  These light particles have a signature imprinted in the early Universe characterized by the effective number of  relativistic species, $N_{\rm{eff}}$.  This can be probed by millimeter-wave CMB observations (\S~\ref{sec_cmb}).
    
\end{itemize}

\section{Current and Near-Future Facilities Relevant to Dark Matter}

This white paper focuses on future facilities that could begin construction and operation in the next decade.  Readers whose primary interest is in those facilities may wish to skip this section.  However, it is useful to set the context by briefly discussing current and near-future facilities that probe dark matter physics. In particular, we note that although the US HEP budget has supported some fraction of the construction and operation for many of these facilities, no funding has been provided for dark matter science effort. Funding to carry out dark matter science with current facilities will be critical for developing the technical and scientific expertise needed for the future experiments that are the focus of this white paper.

\subsection{CMB-S4}
CMB-S4 is a next-generation CMB experiment that received strong support from the Astro2020 Decadal Survey and has achieved CD-0 within the DOE. The CMB-S4 science reach for constraining dark matter and light relics is discussed extensively elsewhere \citep{CMB-S4:2016}, and we leave a further discussion in the context of Snowmass to a white paper from the CMB-S4 Collaboration \citep{CMB-S4-Snowmass}. CMB-S4 will develop critical technical and scientific expertise for future CMB experiments with the potential to constrain dark matter and inflation.

\subsection{Dark Energy Survey (DES)}
The Dark Energy Survey (DES) is a DOE-funded Stage III dark energy experiment using the 4-m Blanco Telescope in Chile. 
DES has delivered impactful measurements of dark matter physics at low and high redshift \citep{Bechtol:2015,Koposov:2015,Drlica-Wagner:2015b,Drlica-Wagner:2015,Chen:2021}. In particular, Milky Way satellite galaxies discovered by DES have enabled strong constraints on the allowed particle mass ranges for thermal relic dark matter and fuzzy dark matter, as well as the dark matter-proton scattering cross section \citep{Nadler:2021}.
However, funding for DES from the US HEP community has focused solely on the dark energy mission objectives.
DES is now in the final analysis stages, and thus is not a topic for the current Snowmass process, but we note that it has been critical for developing scientific expertise for dark matter physics with the upcoming Rubin Observatory.

\subsection{Dark Energy Spectroscopic Instrument (DESI)}
\label{sec:desi}
The Dark Energy Spectroscopic Instrument (DESI) is a Stage~IV dark energy instrument located on the 4-meter Mayall Telescope at Kitt Peak National Observatory. Full survey operations began in 2021 and are supported by the DOE Office of Science. The primary goal of DESI is to constrain the properties of dark energy using baryon acoustic oscillations observed in the three-dimensional clustering of galaxies, quasars, and the Ly-$\alpha$ forest. Although DESI was designed specifically as a dark energy experiment, it presents many avenues for studying dark matter through spectroscopic observations of the nearby and distant Universe.
For more information, we direct the interested reader to the white paper contributed by the DESI Collaboration \citep{DESI-Snowmass}.
We note that dark matter science with DESI is currently unfunded by the US HEP program, but that developing expertise in DESI is extremely relevant to the future dark matter facilities discussed in this paper (see \S~\ref{sec_wfmos}).

\subsection{Rubin Observatory LSST}
Originally conceived as the ``Dark Matter Telescope'' \citep{Tyson:2001}, the Vera C.\ Rubin Observatory Legacy Survey of Space and Time (LSST) provides extensive opportunities as a dark matter facility through observations of gravitational lenses, stars, stellar streams, dwarf galaxies, and large-scale structure \citep{Drlica-Wagner:2019}.  Telescopes such as the Rubin Observatory will also discover a factor of five more Solar System minor objects, which can help constrain the dark matter abundance and ultralight dark matter effects through measurements of their precessions \cite{2020arXiv200907653V,Pitjev:2013,Poddar:2020exe,Tsai:2021irw}. Since the Rubin Observatory is already under construction, we do not to discuss it extensively in this white paper. However, we do emphasize that dark matter science with the Rubin Observatory is currently not funded by the US HEP program.
Funding for dark matter science with Rubin will be critical for developing the scientific expertise to maximize the impact of future dark matter facilities. We leave further discussion of this issue to a white paper from the Dark Matter Working Group within the LSST Dark Energy Science Collaboration \citep{DESC-Snowmass}.

It is worth mentioning that LSST is funded as a 10-year survey program, which is currently scheduled to end in 2034.  The Rubin Observatory as a facility will certainly retain great utility for dark matter science even with 10 years of data in hand.  As of now, no specific plans have been made for the use of the Rubin facility beyond that time horizon.  However, given the capabilities of the telescope, either as constructed or with possible additional upgrades, we encourage the community to develop concepts for future probes of dark matter with the Rubin Observatory.  In particular, we note that operation of the Rubin Observatory beyond its first 10 years offers the potential for increased synergy with ELTs, which may not be completed before the end of the LSST survey. The future of the Rubin Observatory is discussed in more detail in another white paper \cite{Rubin-II-Snowmass}.

\subsection{James Webb Space Telescope}
\label{sec:jwst}
The James Webb Space Telescope (JWST) affords several key measurement opportunities to constrain dark matter properties.  High-redshift observations by JWST could provide constraints on the nature of dark matter based on the abundance of the smallest galaxies \citep{Schultz2014,Dayal2015,Khimey2021}.  Much of this work has focused on articulating differences in early structure formation between warm dark matter and cold dark matter cosmologies.  The dependence on baryonic processes has also been investigated, though there are still significant uncertainties in modeling baryonic processes in galaxies \citep{Khimey2021,Rudakovskyi2021}.  JWST can also contribute significantly to measurements of flux ratio anomalies in gravitationally lensed quasars (see \S~\ref{sec_elts}), which are sensitive to the dark matter halo mass function, because of its high angular resolution and sensitivity at mid-infrared wavelengths.  A complementary route to constrain the nature of dark matter with JWST can also be realized by measuring the accelerations of stars within the Milky Way.  The \textit{Kepler} mission observed  eclipsing binaries \citep{Southworth2015, Windemuth2019} so precisely that the shift in the eclipse mid-point time induced by the Galactic potential in the intervening decade is in principle measurable today with JWST \citep{ChakrabartiET}.  Although the shift in the eclipse mid-point induced by the Galactic potential is very small, $\sim$ 0.1 seconds, space-based missions that afford exquisite photometric and timing precision for eclipsing binary stars can detect it with sources for which other dynamical effects are sub-dominant to the Galactic signal.  For the sample of $\sim200$ eclipsing binaries for which \textit{Kepler} established a very precise baseline a decade earlier, JWST can in principle now provide a new route for direct acceleration measurements and constraints on dark matter sub-structure \cite{ChakrabartiET}.  

\subsection{Nancy Grace Roman Space Telescope}

The \emph{Nancy Grace Roman Space Telescope} (RST) is a 2.4\,m space telescope scheduled to be launched in the mid-2020s by NASA \citep{spergel2015a,akeson2019a}.  RST will have spatial resolution comparable to the \emph{Hubble Space Telescope}, but a $>$100$\times$ larger field of view and higher sensitivity in the near infrared, providing a large increase in survey capabilities.  RST will undertake an ambitious set of mission surveys including a large-area cosmology survey ($\sim2000~\mathrm{deg}^2$) with multiband imaging ($H\sim26.7$\,AB magnitude limit) and slitless grism spectroscopy ($\sim10^{-16}~\mathrm{erg}~\mathrm{s}^{-1}~\mathrm{cm}^{-2}$ line sensitivity at $\lambda\sim1.5$\,$\mu$m).

The cosmology mission surveys are designed to constrain structure formation and the expansion history of the Universe through weak lensing and baryon acoustic oscillations.  However, RST will also make significant contributions to measuring the dark matter halo mass function through strong gravitational lensing. RST and the Rubin Observatory will be the best sources of lensed quasars; Ref. \cite{oguri2010a} predicted that the RST cosmology survey would increase the sample of quadruply-imaged lenses by more than a factor of twenty to $\sim1000$ objects. As described in \S~\ref{sec_elts}, diffraction-limited imaging and spectroscopy of large samples of lenses with ELTs will provide one of the best avenues for pushing mass function measurements to unprecedentedly low masses.  

In addition, RST will be able to map the resolved stellar halos for hundreds of galaxies within $\sim$10\,Mpc, improving current sample sizes by two orders of magnitude.  These maps of the debris that remains after $\sim13$\,Gyr of galaxy accretion and disruption will reveal significant samples of tidal streams \citep{hendel2019,pearson2019} that can be used to constrain the existence of low mass dark matter halos \citep{banik2019}.  Within the Milky Way, RST can enable even more precise measurements of accelerations by measuring the shift in the eclipse mid-point time \citep{ChakrabartiET} since the \textit{Kepler} mission to provide unprecedented constraints on dark matter sub-structure.  While these measurements are possible today with HST and JWST, like all acceleration measurements their precision improves significantly with time, and in this case quadratically with time.  Thus, RST observations a decade in the future will significantly improve the constraints on dark matter sub-structure that are possible today with very precise observations of exquisitely timed eclipsing binary stars.

\subsection{Advanced LIGO}
The gravitational-wave discoveries made by NSF-funded Advanced LIGO detectors and the French-Italian Advanced Virgo detector have revolutionized our view of the universe~\cite{LIGOScientific:2016aoc,LIGOScientific:2017vwq}. The LIGO and Virgo observatories have continued to increase their reach and discovery rate, revealing populations of astrophysical events and routinely issuing alerts to the broader astronomical community. At its 2020 sensitivity, this network was reporting observations of tens of black hole mergers and of order one merger involving a neutron star per year~\cite{LIGOScientific:2021djp,LIGOScientific:2021psn}. In the next few years, the Advanced LIGO and Virgo detectors will continue to observe together, and will be joined by the Japanese KAGRA observatory and by a LIGO detector located in India~\cite{KAGRA:2013rdx}. An upgrade to the Advanced LIGO detectors, known as A$+$, is currently under development and is expected improve the detector's sensitivity by  50\%~\cite{Shoemaker:2019bqt,Miller:2014kma}. These gravitational-wave observatories are exploring the nature of dark matter both directly~\cite{LIGOScientific:2021odm,LIGOScientific:2021job,Nitz:2022ltl}, by the imprint of dark matter on the gravitational-wave signal from merging black holes and neutron stars~\cite{Zhang:2021mks}, and by the effect of dark matter on the masses and spins of observed black holes~\cite{Ng:2020ruv,Ziegler:2020klg}.

\section{Future CMB Facilities}
\label{sec_cmb}

\subsection{Key Measurement Opportunities}

\begin{figure}
\centering
\includegraphics[width=0.47\textwidth]{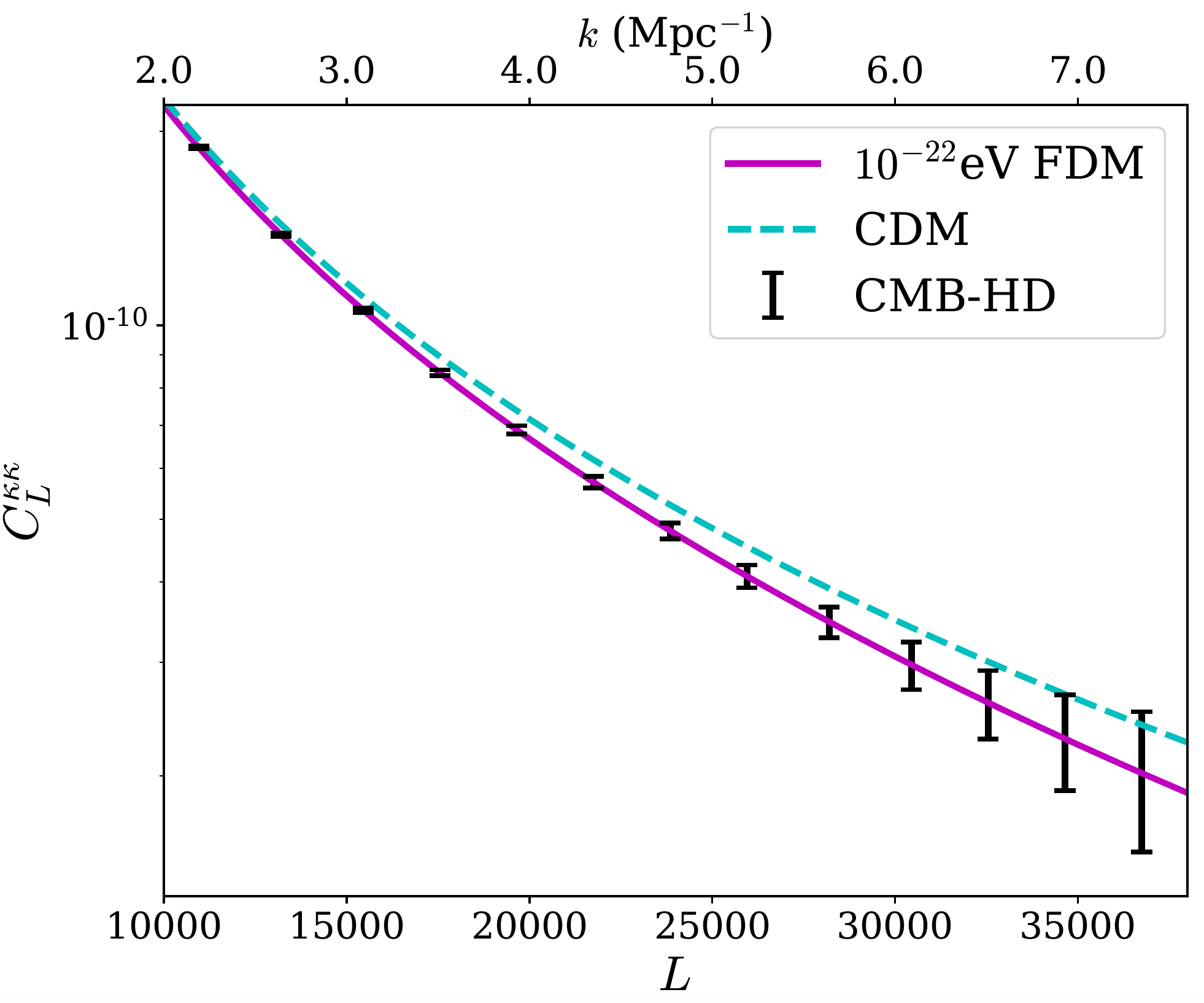}\hspace{4mm}\includegraphics[width=0.5\textwidth,height=6.1cm]{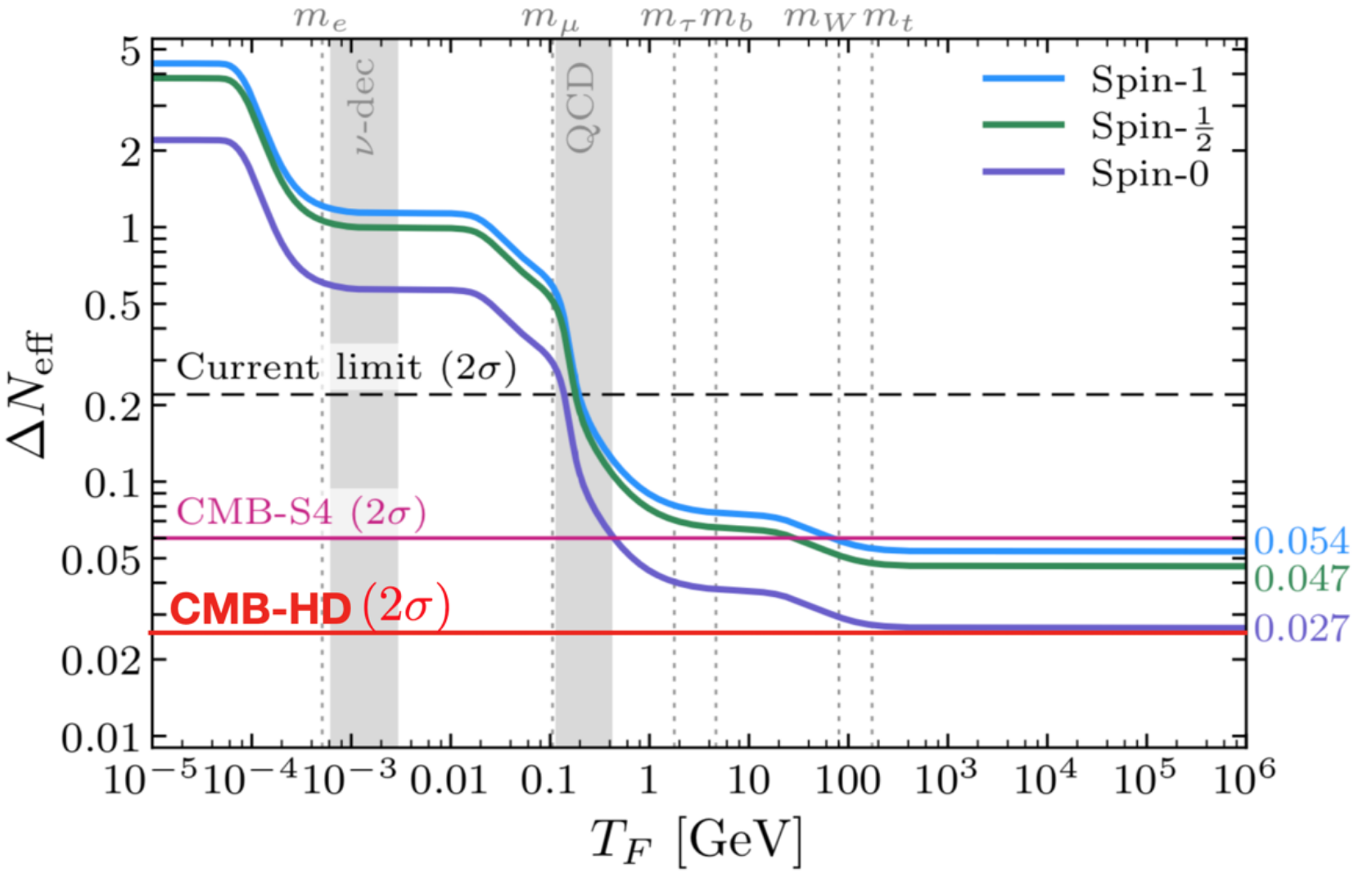}
\caption{{\it{Left:}} CMB-HD would generate via gravitational lensing a high-resolution map, out to $k\sim10~h$Mpc$^{-1}$, of the projected dark matter distribution over half the sky. Shown is the CMB lensing power spectrum for an $m \sim 10^{-22}$ eV FDM model and a CDM model.  Error bars correspond to CMB-HD expectations of 0.5 $\mu$K-arcmin noise in temperature and 15 arcsecond resolution over $50\%$ of the sky.  {\it{Right:}} CMB-HD can achieve $\sigma({N_{\rm{eff}}}) = 0.014$, which would cross the critical threshold of 0.027, ruling out or finding evidence for new light thermal particles, at any time in the early Universe, with at least $95\%$ confidence.  Original figure from~\cite{Green:2019glg,Wallisch:2018rzj}; modified with addition of CMB-HD forecast.}
\label{fig:DMandNeff}  
\end{figure}

To date, some of the most compelling evidence for the existence of non-baryonic dark matter comes from measurements of the Cosmic Microwave Background (CMB) temperature and polarization power spectra, which map the gravitational interplay between dark and baryonic matter in the early Universe.  Mature CMB facilities have allowed us to obtain tight constraints on models of dark matter through the CMB power spectra~\cite{Madhavacheril:2013cna,Planck:2018vyg}, birefringence~\cite{Namikawa:2020:ACT-biref,Bianchini:2020:SPT-biref}, polarization oscillation~\cite{Fedderke:2019ajk,BICEPKeck:2020hhe,BICEPKeck:2021sbt}, and lensing measurements~\cite{Hlozek:2016lzm,Hlozek:2017zzf,Li:2018zdm}.  Continuing this tremendous progress, future CMB measurements have vast potential to constrain or detect dark matter particle properties by exploiting the novel regime that a lower-noise, higher-resolution CMB facility will open. Some of the key measurements anticipated from such a new facility include:

\begin{enumerate}
    \item \textbf{CMB Lensing on Small Scales.} Increasing the resolution and lowering the noise of CMB surveys by at least a factor of five compared to precursor surveys would enable the measurement of the small-scale matter power spectrum on scales of $k\sim10~h$Mpc$^{-1}$ from weak gravitational lensing using the CMB as a backlight. A measurement that would distinguish between the matter power spectrum predicted by cold dark matter (CDM) and that predicted by models that can explain observational puzzles of small-scale structure, with at least $5\sigma$ significance, requires instrument noise levels in temperature of 0.5\,$\mu$K-arcmin over half the sky and a resolution of $\sim15$ arcseconds~\cite{Han:2021vtm,Sehgal:2019nmk,Sehgal:2019ewc} (see left panel of Figure~\ref{fig:DMandNeff}). Such a measurement would result in a high-resolution map of the projected dark matter distribution over half the sky.  The CMB-S4 under construction and precursor CMB experiements can only reach $k\sim1\,h$\,Mpc$^{-1}$ given their sensitivity and noise levels. Using gravitational lensing to measure the matter power spectrum on these scales removes the uncertainty inherent in relying on  baryonic tracers of the matter distribution. The advantages of using the CMB as a backlight to measure gravitational lensing is that 1)~it is at a well-defined redshift, 2)~it has well-known intrinsic properties, and 3)~it is behind all collapsed dark matter structure.  This measurement of small-scale CMB lensing would constrain ultra-light axions (with $m_{a}\simeq 10^{-22}~{\rm eV})$, warm dark matter (with $m_{\nu}\simeq 2~{\rm keV}$), self-interacting dark matter, and any other dark matter model that alters the matter power spectrum on scales of $k\sim10\,h$\,Mpc$^{-1}$~\cite{Markovic:2013iza,Hlozek:2016lzm,Hlozek:2017zzf,Nguyen:2017zqu,Li:2018zdm,Sehgal:2019nmk,Sehgal:2019ewc}.  If the dark sector has multiple components, these measurements could also determine the energy density in a hypothetical sub-dominant component that affects CMB weak lensing, such as structure suppression from ultra-light axions with $10^{-27}~{\rm eV}\lesssim m_{a}\lesssim 10^{-23}~{\rm eV}$. A challenge facing the interpretation of the small-scale dark matter distribution is the degeneracy between suppression of structure due to baryonic effects and suppression due to alternative models of dark matter.  An advantage of measuring the full shape of the CMB lensing power spectrum is that the shape information can potentially be used to distinguish between baryonic effects and dark matter models since each impacts the shape in significantly different ways~\cite{Nguyen:2017zqu}.  
    
    \item \textbf{Number of Light Particle Species.} A CMB survey with lower noise levels would also enable measurements of the number of light particle species, $N_{\rm{eff}}$, that were in thermal equilibrium with the known standard-model particles at any time in the early Universe, with a $1\sigma$ uncertainty of $\sigma({N_{\rm{eff}}}) = 0.014$. This accuracy would cross the critical threshold of 0.027, which is the amount that any new thermal particle species changes $N_{\rm{eff}}$ away from its Standard Model value of 3.04.  In contrast, the expected target of CMB-S4 is $\sigma({N_{\rm{eff}}}) = 0.03$~\cite{Abazajian:2019eic}. Achieving $\sigma({N_{\rm{eff}}}) = 0.014$ would rule out or find evidence for new light thermal particles with at least $95\%$ confidence, and is possible with instrument noise levels of 0.5\,$\mu$K-arcmin over half the sky if the measurements are made at high enough angular resolution ($\sim15$ arcseconds) to remove contaminating foregrounds~\cite{Sehgal:2019nmk,Sehgal:2019ewc} (see right panel of Figure~\ref{fig:DMandNeff}). Such a measurement is important, since many dark-sector models predict additional new light thermal particle species~\cite{Baumann:2015rya,Boehm:2013jpa}.
    
    \item \textbf{CMB Photon Conversion to Axion-like Particles.} CMB surveys can also constrain or discover axion-like particles by observing the resonant conversion of CMB photons into axion-like particles in the magnetic fields of galaxy clusters. Nearly massless pseudoscalar bosons, often generically called axion-like particles, appear in many extensions of the standard model~\cite{PhysRevLett.38.1440,Weinberg:1977ma,PhysRevLett.40.279,Svrcek:2006yi,Arvanitaki:2009fg,Acharya:2010zx}. A detection of such particles would have major implications both for particle physics and for cosmology, not least because axion-like particles are also a well-motivated dark matter candidate. A CMB experiment with 0.5\,$\mu$K-arcmin over half the sky and $\sim15$ arcsecond resolution would provide a world-leading probe of the electromagnetic interaction between axion-like particles and photons using the resonant conversion of CMB photons and axion-like particles~\cite{Raffelt:1996wa,Mukherjee:2018oeb} in the magnetic field of galaxy clusters~\cite{Mukherjee:2019dsu}.  It would explore the mass range of $10^{-13}$\,eV~$<m_a \lesssim 2\times 10^{-12}$\,eV and improve the constraint on the axion coupling constant by over two orders of magnitude over current particle physics constraints to $g_{ a \gamma}<0.1\times 10^{-12}$\,GeV$^{-1}$. These ranges are unexplored to date and are complementary with other cosmological searches for the imprints of axion-like particles on the cosmic density field. 
    
    \item \textbf{Time-dependent CMB Polarization Rotation.} Another feature of CMB surveys is that they can constrain or discover axion-like dark matter by measuring time-dependent CMB polarization rotation. Ultralight axion-like dark-matter fields that couple to photons via $g_{a\gamma}$ cause a time-dependent photon birefringence effect which manifests as a temporal oscillation of the local CMB polarization angle (i.e., a local $Q\leftrightarrow U$ oscillation in time)~\cite{Fedderke:2019ajk}. This rotation effect is in-phase across the entire sky, and the oscillation period is fixed by the axion-like particle mass to be at observable timescales of $\sim$months to $\sim$hours for masses in the range $10^{-21}\,\textrm{eV}\lesssim m_a \lesssim 10^{-18}\,\textrm{eV}$. Searches for this effect are not limited by cosmic variance since it is a time-dependent oscillation of the \emph{observed} CMB polarization pattern. Improvements in the polarization map-depth and sky coverage achieved by reaching noise levels of 0.7 $\mu$K-arcmin in polarization over half the sky promise improvements for this search by a couple of orders of magnitude in the coupling $g_{a\gamma}$ as compared to existing BICEP/Keck analyses~\cite{BICEPKeck:2020hhe,BICEPKeck:2021sbt}, and exceed the CMB-S4 projected reach by a factor of $\mathcal{O}(2)$.
    
    \item \textbf{Cosmic Birefringence.} The novel low-noise regime of future CMB surveys can enable world-leading constraints on beyond standard model physics.  In particular, such surveys enable measurements of isotropic birefringence and the scale-invariant birefringence power spectrum to test for new physics by searching for a rotation of linear polarization as the CMB photons propagate to us from the surface of last scattering.  This polarization rotation, called cosmic birefringence, can be used to constrain very light axion-like particles of $m_a \lesssim 10^{-28}$\,eV \cite{Harari:1992:axion,Carroll:1998,Li:2008,Pospelov:2009,Capparelli:2019:CB}, the axion string network \cite{Agrawal:2019:biref}, axion dark matter \cite{Liu:2016dcg}, general Lorentz-violating physics in the context of Standard Model extensions \cite{Leon:2017}, and primordial magnetic fields through Faraday rotation \cite{Kosowsky:1996,Harari:1997,Kosowsky:2004:FR}. Achieving 0.7 $\mu$K-arcmin in polarization over half the sky will result in $1\sigma$ uncertainties of 0.035 arcmin and $1.4 \times 10^{-6}$ deg$^2$, for isotropic birefringence and the scale-invariant birefringence power spectrum, respectively. This will improve statistical uncertainties on birefringence measurements by several orders of magnitude over the existing constraints \cite{Planck:2016soo,Choi:2020ccd,Minami:2020:biref,Namikawa:2020:ACT-biref,Bianchini:2020:SPT-biref,Mandal:2022tqu}.
\end{enumerate}    

\subsection{CMB-HD}
\label{sec:cmb_hd}

CMB-HD is an ambitious leap beyond previous (e.g., SPT-3G, ACT) and upcoming (e.g., Simons Observatory, South Pole Observatory, CMB-S4) ground-based millimeter-wave experiments. CMB-HD consists of two new 30-meter-class off-axis crossed Dragone telescopes to be located at Cerro Toco in the Atacama Desert.  Each telescope would host 800,000 detectors (200,000 pixels), for a total of 1.6 million detectors. The CMB-HD survey would cover half the sky ($\sim$20,000\,deg$^{2}$) over 7.5 years.  These observations would result in an ultra-deep, high-resolution millimeter-wave map with 0.5 $\mu$K-arcmin instrument noise in temperature (0.7\,$\mu$K-arcmin in polarization) in combined 90 and 150\,GHz channels and 15-arcsecond resolution at 150\,GHz.  For comparison, this survey would have three times lower noise and six times higher resolution than the CMB-S4 experiment~\cite{Abazajian:2019eic}, opening a qualitatively new regime of millimeter-wave science.  CMB-HD would also observe at seven different frequencies between 30 and 350\,GHz to mitigate foreground contamination. 

Recently, the Astro2020 Decadal Survey report expressed strong support for CMB science in general, and indicated that it had a long future horizon of science to offer.  CMB-HD would cross important thresholds and provide definitive answers to many fundamental physics questions, including the nature of dark matter, the light particle content of the Universe, the existence and mechanism of inflation, and whether there is new physics in the early Universe beyond the Standard Model, as suggested by recent H$_{0}$ measurements~\cite{Riess:2021jrx}. A summary of the key science goals motivating the CMB-HD survey and the flowdown to measurement and instrument requirements are given in the Astro2020 Science White Paper~\cite{Sehgal:2019nmk}, Astro2020 CMB-HD APC~\cite{Sehgal:2019ewc}, Astro2020 CMB-HD RFI~\cite{Sehgal:2020yja}, and the Snowmass2021 CMB-HD White Paper~\cite{CMB-HD-Snowmass}. In addition, further information can be found at~\href{https://cmb-hd.org}{https://cmb-hd.org}.

\section{Future Pulsar Timing Facilities}
\label{sec:pulsartiming}

\subsection{Key Measurement Opportunities}

 The accelerations of stars provide the most direct window into the mass distributions of galaxies. 
 The motions of stars trace the distribution of gravitating matter and can be used to measure both the smooth and clumpy distribution of dark matter.  For more than a century, these accelerations have been \emph{estimated} by measuring the positions and velocities of stars \citep{Oort1932,Kuijken_Gilmore1989,Bovy_Tremaine2012,McKee2015,Schutz2018}.  Such analyses assume equilibrium in order to relate the phase-space distribution of stars to the potential.  
The large-scale potential is then described by parameters like the average mid-plane density (also known as the Oort limit), the local dark matter density (essential for interpreting direct detection experiments), the shape of the Galactic
potential, and the slope of the rotation curve.  However, observations of the gas and the stellar disk show that our Galaxy has had a highly dynamic history \citep{LevineBlitzHeiles2006, Xu2015,antoja2018, Helmi2018, Belokurov2018}.  In such a time-dependent potential,
direct measurement of accelerations can capture the complexity of the time-dependent Galactic mass distribution, including both the dark matter and stellar contributions, without making any of the assumptions of equilibrium or symmetry that are inherent in kinematic analyses.  Simulations of interacting galaxies find differences in the \emph{true} density in the simulation relative to what is \emph{estimated} from the Jeans analysis by up to a factor of $\sim2$ in the solar neighborhood \citep{Haines2019}, which shows that the assumption of equilibrium for a time-dependent potential like that of the Milky Way does not produce accurate results in characterizing the mass density of the Galaxy, or other fundamental Galactic parameters.  Using the Poisson equation, $\nabla^2 \Phi = - \nabla \cdot \vec a = 4 \pi G \rho $, acceleration measurements can be straightforwardly related to the Galactic potential, $\Phi$, and the mass density, $\rho$, without assumptions.

Stellar accelerations in the Milky Way are small (of order $\sim10$~cm~s$^{-1}$ over a decade, or about $\sim 3 \times 10^{-8}~\rm cm~s^{-2}$, for stars within $\sim$ a few kpc of the Sun), but these measurements have recently become possible with analyses of compiled pulsar timing observations \citep{Chakrabarti2021}.  Ref.~\cite{Chakrabarti2021} used the observed time-rate of change of the binary period, $\dot{P}_{b}$, of fourteen binary millisecond pulsars to derive the Galactic acceleration.  The best-timed  pulsars have an extent of $\sim1$~kpc in radius and vertical height relative to the Sun, and therefore provide a more significant constraint in the smaller dimension, i.e., the vertical dimension.  Given the pulsar accelerations, the Oort limit was measured to 3$\sigma$ significance.  The local dark matter density determined from pulsar timing currently has large errors, but should improve significantly with future pulsar timing facilities. The shape of the Galactic potential traced by the pulsars was also tightly constrained, and resembles a disk rather than a spherical halo \citep{Chakrabarti2021}. These observations thus far come from current radio facilities, including the North American Nanohertz Observatory for Gravitational Waves (NANOGrav) \cite{Nano2018}, the European Pulsar Timing Array (EPTA) \citep{EPTA2016}, and the Parkes Pulsar Timing Array (PPTA) \cite{PPTA}.
 
Future radio telescopes like ngVLA and DSA-2000 will enable more sensitive pulsar timing observations that have the potential to carry out precise measurements out to larger distances in the Galaxy. In addition, these data will constrain dark matter sub-structure down to $\sim 10^{6}$\,M$_{\odot}$, and could thereby eliminate competing dark matter models.  These facilities will benefit from both improvements in increased sensitivity and longer time baseline observations.  Very roughly (i.e., for a homogeneous dataset), the improvement in precision for $\dot{P}_{b} \propto t^{-5/2} SNR^{-1}$ \citep{LorimerKramer2012}.  The primary limiting factor for acceleration measurements presently (and derived properties such as the local dark matter density) is the precision of the observed $\dot{P}_{b}$ and parallaxes as determined from pulsar timing.  These measurements will improve significantly as noted above for a homogeneous dataset, although in practice  how well different components of the acceleration are recovered will depend on the distribution of sources. The point-source sensitivity for ngVLA is expected to be about ten times better than that of VLA or GBT \citep{Selina2018}.  Observations enabled by ngVLA are expected to reach sufficient precision for the best-timed pulsars to produce acceleration measurements out to the Magellanic Clouds, and possibly beyond \citep{Nanograv}.  Fluctuations in the gravitational potential along the line of sight to radio pulsars, such as might arise from clumps of dark matter or primordial black holes, can be discerned by correlating or comparing different lines of sight.  ngVLA would deliver higher precision pulsar timing by virtue of its higher sensitivity than existing telescopes, particularly at higher frequencies ($\sim$ 3 GHz), which are used to mitigate systematic effects due to the propagation through the Galaxy’s interstellar medium.

Techniques to characterize dark matter using pulsar timing observations are reviewed in Ref. \citep{BurkeSpolaor2019}, and include the possible detection of very small-scale dark matter clumps from the Doppler delay due to the acceleration of a pulsar (or the Earth) from an intervening compact dark matter clump. 
The Square Kilometer Array (SKA) is expected to provide sensitivities to $\sim 10^{-9}$~M$_{\odot}$ compact sub-halos, assuming concentrations of $c \sim 10^8$ and extremely close passages, of order $\sim 10^{-3}~\rm pc$ \citep{Dror2019}.  This work was later generalized to a passing ensemble of sub-halos \citep{Ramani2020}.  One potential caveat in determining accelerations from spin periods arises from the well-known degeneracy between the observed spin-down of a pulsar and magnetic braking \citep{LorimerKramer2012}.  An especially interesting constraint on the nature of dark matter that can realized by next-generation pulsar timing facilities is the detection of periodic variations in the pulse arrival times, as would be produced by ultralight scalar field dark matter which leads to an oscillating gravitational potential at nanohertz frequencies.  The Parkes Pulsar Timing Array searched for this signal but did not find a statistically significant detection \citep{Porayko2018}; future facilities should improve these searches in the regime of astrophysically allowed masses ($\gtrsim 10^{-22}~\rm eV$).

\subsection{ngVLA \& DSA-2000}
\label{sec:ngvla_dsa2000}

The next-generation Very Large Array (ngVLA) is an interferometric array that will provide an order-of-magnitude improvement in sensitivity relative to the Jansky VLA and ALMA operating at the same wavelengths, as well as factor of 30 increase in baselines, yielding milliarcsecond resolution \citep{Selina2018}. The ngVLA operates at frequencies of 1.2\,GHz (25\,cm) to 116\,GHz (2.6\,mm), building on the legacy of the Jansky VLA, ALMA and the VLBA as the next major national facility in ground-based radio astronomy \citep{Selina2018}. This proposed radio interferometer is optimized to bridge the gap between ALMA at wavelengths of 2\,mm and shorter, and the future Square Kilometer Array (SKA) at decimeter to meter wavelengths. The ngVLA opens a new window on the Universe through ultra-sensitive imaging of thermal line and continuum emission down to milliarcsecond resolution, while also delivering unprecedented broadband continuum imaging and polarimetry of non-thermal emission. 

The ngVLA will comprise an array of 214 antennas, each 18\,m in diameter, supplemented with a short baseline array of 19 smaller 6\,m-diameter antennas, to deliver the shortest baselines (largest structures on the sky). Each fixed antenna is outfitted with receivers spanning the frequency range 1.2--50.5\,GHz (25\,cm--5.9\,mm) and 70–-116\,GHz (4.3\,mm--2.6\,mm). The array achieves high surface brightness sensitivity and high-fidelity imaging on angular scales down to 1\,mas, by distributing a large fraction of the total collecting area in a core array area with extensions out to $\sim$1000\,km. An additional 30 18\,m antennas are placed in stations to provide even longer baselines (up to 8,860\,km) reaching across North America and Hawai'i. These antennas can be used separately or together with the main array, and will deliver 0.1\,mas resolution and enable microarcsecond precision interferometric astrometry. The ngVLA was highly ranked by the Astro2020 Decadal Survey, it is currently beginning the construction and testing of a prototype antenna, and can potentially start full construction by mid-decade. The ngVLA can operate both in interferometric and phased-array modes, and can be broken in up into subarrays. One of the ngVLA Key Science Goals is the study of pulsars in the Galactic Center, and its signal chain is designed from the start to provide up to 10 phased-array beams within the $48^\prime\,(1.2\,{\rm GHz}/\nu)$ antenna field of view, appropriate for pulsar timing and/or searches. Each of these phased-array beams will have a robustly-weighted angular resolution near 20\,mas in Band 3 (12--20~GHz) to 160\,mas in Band 1 (1.2--3.5~GHz).

The Deep Synoptic Array-2000 (DSA-2000) is a proposed radio survey telescope that will consist of 2000 $\times$ 5\,m dishes, having an equivalent point source sensitivity to SKA-mid \citep{Hallinan2019},  spanning an area of 19\,km $\times$ 15\,km in Nevada. The DSA-2000 will instantaneously cover the frequency range from 0.7–-2\,GHz (42--15\,cm). It is designed to serve as an all-sky survey instrument, and is therefore very complementary to ngVLA, and is the radio counterpart to the Rubin Observatory \citep{Hallinan2019}.  It will have near complete sampling of the uv-plane, thereby replacing a traditional correlator digital backend with a “radio camera.”  In a five-year prime phase, the DSA-2000 will image the entire viewable sky ($\sim$30,000 deg$^2$) repeatedly over sixteen epochs, detecting $>$1 billion radio sources in a combined full-Stokes sky map with 500\,nJy\,beam$^{-1}$ rms noise.  One of the primary objectives of DSA-2000 is to serve as a pulsar timing array (25\% of the time), and it will be the main instrument comprising NANOGrav in the future.  It is expected to produce timing precision better than $\sim1$\,ms for 2000--10,000 millisecond pulsars \citep{Hallinan2019}.

Given measurements of accelerations enabled by future pulsar timing facilities, one can expect to characterize the Galactic potential and fundamental parameters that describe the Galaxy out to tens of kpc.  Constraints on these parameters have been obtained from a variety of analyses of current and prior surveys from kinematic modeling.  Key large-scale descriptors of the Galactic potential include the density profile, the rotation curve, and the shape of the Galactic potential.  Given the improved sensitivity of ng-VLA and DSA-2000, it should be possible to derive these fundamental Galactic parameters from direct acceleration measurements out to large distances.  Relative to earlier determinations of the local dark matter density from pulsar timing \citep{Chakrabarti2021}, one can expect improvements in precision that should be formally competitive with kinematic estimates.  Perhaps, the most exciting aspect for dark matter studies is the ability to constrain not only the smooth component of the mass distribution, but also directly measure dark matter sub-structure, potentially down to the $\sim10^{6}$\,M$_{\odot}$ scale.

\section{Future 21cm Facilities}
\label{sec_21}

\subsection{Key Measurement Opportunities}
The baryonic content of the universe is dominated by hydrogen, which constitutes approximately three quarters of all baryons. Hydrogen recombination produces the CMB and hydrogen is the fuel of star formation. During the different ages and places in the universe, hydrogen can be found in many forms: fully ionized, atomic and molecular. Atomic hydrogen shines in the 21\,cm transition. This is a forbidden very low energy transition between the atomic hydrogen with proton and electron spins being aligned and anti-aligned. It corresponds to a line at frequency of approximately 1420\,MHz, or equivalently, a wavelength of 21\,cm. 

During the evolution of the universe, the dominant emission mechanism through which 21\,cm photons are produced varies between cosmic epochs. These different regimes therefore present different opportunities for constraining dark matter through observations of the 21\,cm signal. They are sensitive to different physics, but the techniques share many technicalities in common and therefore from the instrumentation and data analysis point of view, the problems are variations on a few fundamental issues. The primary challenge at all frequencies relevant for 21\,cm observations is foreground emission, emitted largely as synchrotron radiation from the MW and other galaxies. This foreground emission is spectrally smooth, but to successfully filter it out requires 
instrumental dynamic range spanning 4--8 orders of magnitude. This requires an exquisite instrumental calibration that has never been fully achieved in practice. We emphasize that these are purely technical challenges that are currently being solved by the radio astronomy community. DOE HEP can strongly contribute to his effort through funding the appropriate instrumentation development required to enable these measurements as a part of generic detector development, as well as pursuing full scale system simulations that will inform the next generation of system design and data reduction algorithms. An extended discussion of this technique in the context of DOE HEP, can be found in the dedicated Snowmass2021 White Paper \cite{CF5_21cm}.

\begin{enumerate}
    \item \textbf{Dark Ages Cosmology.} After recombination, the universe was filled with neutral hydrogen and CMB photons. The physics of this universe remains perfectly describable by linearized general relativity, thermodynamics and atomic physics. This remains true until the first stars turn on, after which the complex and highly non-linear physics of star formation prevents us from making first-principle calculations. The so-called Dark Ages, between redshifts $z\sim 1100$ and $z\sim 30$, thus offer a unique cosmological window. The physics is as accurately predictable as that of the CMB, but the field itself is three-dimensional, having 10 orders of magnitude more linear fluctuation modes accessible to study through their 21\,cm emission.  This signal has the potential to enable the ultimate sensitivity to cosmological parameters, including those from early-universe inflation and primordial gravitational waves. Although this experimental end-game will take decades to develop, the first steps toward this eventual goal can be taken now. We note that the Astro2020 Decadal Survey identified Dark Ages Cosmology as the sole Discovery Area from the Panel on Cosmology. 
    
    Although the signal from density fluctuations in the Dark Ages is a ways away, detecting the overall monopole signal --- the Dark Ages equivalent of the discovery of the CMB \citep{PenziasWilson1965} --- is within reach this decade. The monopole in this case corresponds to a net absorption of CMB photons, appearing as a broad spectral dip centered at $\sim20$\,MHz. The shape of this dip is very sensitive to any non-standard physics or dark matter interactions at that epoch and in particular any energy injection (or sink) from the baryon fluid at that epoch. For this reason, the Dark Ages monopole has also been dubbed the \emph{Cosmic Calorimeter}. Because these predictions are naturally robust, deviation would be a smoking gun for new physics. In fact, a large number of theoretical proposals exist that can cause deviations, including dark matter-baryon scattering~\cite{1509.00029,1408.2571,1803.06698,1803.09734}, millicharged dark matter~\cite{1802.10094,1803.03091}, dark-matter annihilation~\cite{1211.0283,1604.02457}, primordial black holes~\cite{1803.09390}, axions~\cite{1804.10378}, neutrino decay, charge sequestration~\cite{1803.10096}, quark nuggets, dark photons~\cite{1809.01139}, interacting dark energy~\cite{1803.07555} and even a drift in the value of the fundamental constants~\cite{astro-ph/0701752}. More details can be found in an Astro2020 Decadal Survey submission \cite{1903.06212}.
    
    Unfortunately, Dark Ages cosmology is not feasible from the Earth because of ubiquitous sources of radio interference, both human-made and meteorological, as well as ionospheric effects that become very difficult to control with precision below 50~MHz. 
    
    \item \textbf{Cosmic Dawn and the Epoch of Reionization.} The formation of the first luminous objects marks the end of the Dark Ages, ushering in the subsequent era of Cosmic Dawn (when Population III stars began to form) and Epoch of Reionization (EoR; when first-generation galaxies formed). This has several important consequences on the behavior of the 21cm line. First, Ly$\alpha$ photons from the first stars induce the Wouthuysen-Field effect, which causes the 21cm brightness temperature to be coupled to the Ly$\alpha$ background. Thus, 21cm emission ceases to simply trace the matter distribution but is instead sensitive to Ly$\alpha$ fluctuations. Eventually, the Ly$\alpha$ background becomes sufficiently strong that it serves as a uniform background since its effects saturate everywhere. Following this, X-ray heating (e.g., from X-ray binaries forming inside the first galaxies) causes temperature variations in the IGM, resulting in spin temperature fluctuations that imprint themselves on the 21cm line. Eventually, these heating effects also saturate, with their effect again becoming that of a uniform background. This cycle repeats itself a final time with reionization, where UV photons from the first galaxies ionize the IGM around them. This has the effect of imprinting strong spatial fluctuations in the form of ionized bubbles, until the entire IGM ionizes.
    
    Observations of the 21\,cm line during Cosmic Dawn and the EoR will be powerful probes of the astrophysics of the era. However, indirect probes of cosmology and fundamental physics still hold considerable promise. For example, a deep (and empirically confirmed) understanding of reionization removes a key nuisance for CMB experiments seeking to constrain fundamental parameters such as neutrino masses \cite{Liu:2016,Billings:2021,Fialkov:2016}. Relative velocity effects between dark matter and baryons \cite{Tseliakhovich:2010} also imprint themselves on the Cosmic Dawn power spectrum, resulting in the fairly model-independent signature of velocity-induced acoustic oscillations (VAOs) \cite{Munoz2019}. These VAOs can be used, for instance, as standard rulers that enable high-redshift determinations of the Hubble parameter \cite{Munoz2019ruler}. Finally, recent studies have suggested that the 21\,cm brightness temperature may be perturbative for at least part of the EoR \cite{McQuinn:2018,Hoffmann:2019}. Thus, there remains the possibility that the astrophysics of the EoR can be brought under control by the marginalization of a small handful of effective parameters, leaving just constraints on the matter field.
    
    \item \textbf{Galaxy Surveys in 21\,cm}. Below redshift of around $z \sim 6$, the intergalactic medium becomes essentially fully ionized. Galaxies produce a sufficient number of ionizing photons that any hydrogen atoms whose density is below that required for self-shielding are maintained in an ionized state. The main exceptions are pockets of dense neutral hydrogen inside individual galaxies. Therefore, it is in principle possible to perform a galaxy survey in 21\,cm. However, this emission is weak and therefore looking at individual galaxies is inefficient. Instead, relying on the fact that there is no important interloper for the 21\,cm line, we can rely on the intensity mapping technique. In this approach, rather than imaging individual galaxies, a low-resolution instrument measures the fluctuations in the mean intensity of 21\,cm emission in space and time. This emission tracks the 21\,cm weighted number density of galaxies and contains all the necessary information required to model the growth of large scale structure. Such surveys are complementary to optical galaxy surveys in many respects. First, the signal is dominated by galaxies residing in considerably less massive dark matter halos than those typically targeted with optical galaxy surveys, with the majority of the signal coming from dark matter halos of mass $\sim 10^{10} M_\odot$. The main advantage of such a survey is that it is almost invariably limited by the thermal noise of the receiver rather than Poisson noise of the underlying sources. By probing a large number of linear modes, this epoch enables exceptionally tight constraints on a range of postulated dark matter candidates \cite{Gluscevic19}.  Some of these constraints follow indirectly from measuring the radiation content of the universe, i.e. the $N_{\rm eff}$ parameter \cite{Green17,Green21}, but other models provide a more direct route to large changes in observable correlations.  Examples of non-standard models probed by high-precision power spectrum measurements include those with dark matter-baryon scattering \cite{Dvorkin14}, dark matter interactions \cite{Leagourgues16,Pan18,Archidiacono19} or ultra-light axions \cite{Hlozek15}. Moreover, a very broad class of DM models that can be described with the effective theory of structure formation (ETHOS) framework \cite{ETHOS}.
\end{enumerate}

\subsection{LuSEE-Night}
\label{sec:lusee_night}

\noindent LuSEE (Lunar Surface Electromagnetics Experiment) Night is a proposed collaboration between DOE and NASA to install a test-bed radio receiver on the far side of the Moon. It will be launched on the Commercial Lunar Payload Services flight CS-3 in early 2025. It will consist of 4 monopole antennas that actuate in altitude and azimuth. The lander and all the other experiments will become electrically inactive during the first lunar dawn and LuSEE Night will employ strong mitigation techniques to prevent self-contamination with radio frequency interference. The current plans include a far-field calibrator on the satellite that will deliver a coded signal from orbit and enable precise characterization of instrumental response, including that of the lunar regolith. LuSEE Night is expected to acquire data for at least 12 lunar nights. At the end of the mission it will provide the most exquisite measurements of the low-frequency radio sky below 50~MHz to date and demonstrate the feasibility of Dark Ages cosmology from the far side of the Moon.

\subsection{PUMA}
\label{sec:puma}
\noindent The Packed Ultra-wideband Mapping Array (PUMA) is a proposed wideband radio instrument to map half the sky in the 200--1100\,MHz frequency range, corresponding to the redshift range $z\sim 0.3 - 6$ when used as a 21\,cm intensity mapper. PUMA is a very large transit radio array proposed in two configurations. A 5000- dish configuration (PUMA-5K) can act as a pathfinder while still completing important science goals, including very high precision measurements of the expansion history of the universe, while the full-sized array will approach sample-variance-limited observations.  As discussed above, PUMA will provide strong constraints on dark matter properties in particular through high precision measurement of the 2-point and 3-point correlation function at high redshift. Its sample-variance-limited measurement at lower redshift could open up new avenues for dark matter studies, but the details remain open for further theoretical investigation.

\section{Future Optical Facilities}

\subsection{Key Measurement Opportunities}
\label{sec:optical_measurement_opportunities}

Many observations that focus on detecting the effects of dark matter via its gravitational interactions with baryons have been made at optical wavelengths due to the combination of telescope sensitivity and the fact that the energy output from stars peaks at these wavelengths.  Dark matter is studied with optical telescopes in two primary ways: using the observed velocities and accelerations of stars, gas, and galaxies to infer the gravitational potential that is causing their motions, and using gravitational lensing to calculate the mass surface density along the line of sight to the lensed object. 
Despite the relative maturity of this observational domain, future instruments and facilities are poised to make significant technological advances in image resolution, sensitivity, and the number and precision of spectroscopic velocity measurements.
These advances will greatly increase the observational sensitivity to fundamental dark matter properties.
Some of the key measurements that will be enabled by new facilities include:

\begin{enumerate}
    \item The mass function of dark matter halos from $10^{6}-10^{8}$~M$_{\odot}$.  The upper end of this range will be probed directly by ELT measurements of the masses of dwarf galaxies that will be discovered by the Vera Rubin Observatory \citep[e.g.,][]{Nadler2020}.  Dark matter halos below $10^{7}$~M$_{\odot}$ are unlikely to contain any baryons, and therefore cannot be seen directly.  These halos can be detected in the Milky Way from the perturbations they impart to dynamically cold stellar streams \citep[e.g.,][]{Ngan2014,Erkal2015,Bonaca2019,Drlica-Wagner:2019} and the Galactic acceleration field \citep{Chakrabarti2020}, and at cosmological distances from the changes in image brightnesses and positions they cause in gravitational lens systems \citep[e.g.,][]{Vegetti2012,Hezaveh2016,Gilman2020}.  The former can be measured by wide-field spectroscopic facilities (e.g., MegaMapper, MSE, and SpecTel) and extreme-precision radial velocity spectrographs, and the latter by integral field spectrographs on ELTs.  The halo mass function is closely tied to the primordial matter power spectrum and fundamental dark matter properties such as particle mass.  The mass function also constrains interactions between dark matter and standard model species \citep[e.g.,][]{Nadler19,Maamari21,Nguyen21}.
    
    \item The central densities of the smallest luminous dark matter halos ($10^{8}-10^{9}$~M$_{\odot}$).  In small dwarf galaxies, the measured stellar velocity dispersion translates directly to the mean density of dark matter in the innermost $\sim100$~pc of the galaxy \citep[e.g.,][]{Read19,Guerra2021}.  For the closest Milky Way satellites, these stellar kinematics can be obtained with existing instruments, but for more distant systems that have not been altered by interactions with a massive host galaxy, ELTs will be needed.  The maximum density dark matter can reach on these scales is affected by the dark matter particle mass and its self-interaction cross-section at low velocities.
    
    \item The internal mass distribution of dark matter halos in dwarf galaxies.  The standard $\Lambda$CDM model makes clear predictions for halo density profiles \citep{NFW1996}, which largely do not match those determined observationally \citep[e.g.,][]{Oh2011,Adams2014,Relatores2019}.  A cold, collisionless particle should result in a cuspy ($\rho(r) \propto r^{-1}$) inner profile, with a steeper decline at large radii, although the degree to which baryons should alter the pure dark matter expectation as a function of halo mass remains controversial \citep[e.g.,][]{Orkney2014,Wheeler2019,Lazar2020}.  Robust measurements of the density profiles in the most dark matter-dominated dwarf galaxies are a highly-anticipated goal of next-generation facilities.  Two observational approaches are possible: radial velocities for $\sim10^{4}$ stars per dwarf galaxy from wide-field spectroscopic facilities, or proper motions for $\sim10^{3}$ stars per dwarf from ELTs \citep{Guerra2021}.  
    
    \item The mass and density profile of the Milky Way.  The total mass of our Galaxy is a crucial variable for understanding the results of both astrophysical dark matter measurements and direct detection experiments.  Currently uncertain at the factor of $\sim2$ level \citep[e.g.,][]{Deason2021,Craig2021,Necib2022}, the Milky Way mass can be determined with percent-level accuracy with future measurements of velocities in stellar streams by wide-field spectroscopic facilities \citep{Bonaca2018}.  The velocities of stars in the outer reaches of the Milky Way's dark matter halo, obtained with the same instruments, may also contribute to this measurement.  The Milky Way mass determines the expected abundance of satellite dark matter halos (see above) as well as the velocity distribution of dark matter particles near the Earth.  Galactic acceleration measurements from extremely precise time-series radial velocities and eclipse timing (\S~\ref{sec:jwst}) can also improve constraints on the overall mass distribution within the Milky Way, providing dark matter limits as described in point 3 above.
    
    \item The density of dark matter in the solar neighborhood can be determined in less than a decade with current-generation high-precision spectrographs (with RV precision $\sim10$~cm~s$^{-1}$) from direct acceleration measurements \citep{Chakrabarti2020}, and compared to earlier determinations from pulsar timing.  This quantity is critical for the interpretation of dark matter direct detection experiments.
    
\end{enumerate}

\begin{figure*}
\includegraphics[width=0.49\textwidth]{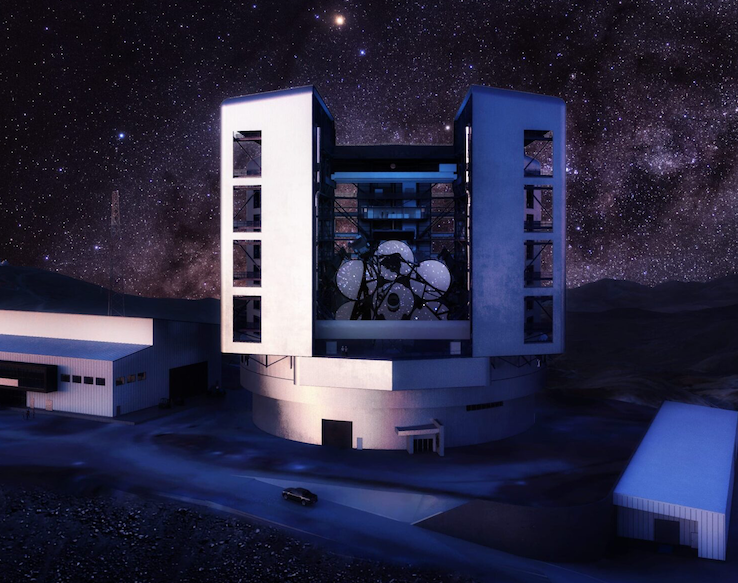}
\includegraphics[width=0.49\textwidth]{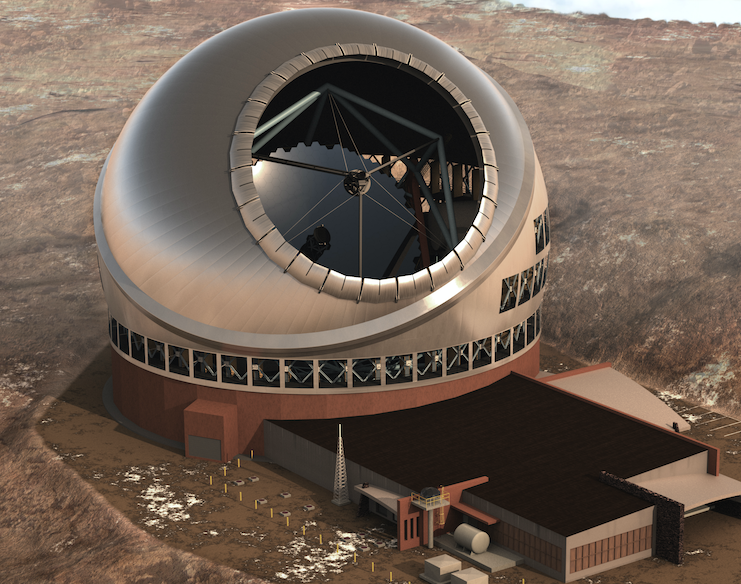}
\caption{\label{fig:elt} Conceptual designs of the Thirty Meter Telescope (left; courtesy TMT International Observatory) and the Giant Magellan Telescope (right;  Giant Magellan Telescope – GMTO Corporation). Both projects are at an advanced stage of development and have commenced construction of their primary optics.}
\end{figure*}

\subsection{Extremely Large Telescopes}
\label{sec_elts}

Two US-led partnerships are in the process of building giant ground-based telescopes with apertures of $25-30$~m, offering an order of magnitude increase in collecting area and a factor of three improvement in the diffraction limit compared to the largest current optical telescopes.  The Giant Magellan Telescope (GMT) is a 24.5~m telescope under construction at Las Campanas Observatory in Chile, and the Thirty Meter Telescope (TMT) is a 30~m segmented mirror telescope planned to be located in either Hawaii or the Canary Islands.  The highest priority recommendation from the Astro2020 decadal survey for ground-based facilities is that the federal government join both projects as the dominant partner to provide access to extremely large telescopes (ELTs) for the entire US astrophysics community.

ELTs are expected to make significant contributions to dark matter science through observations of gravitational lenses, dwarf galaxies, and direct acceleration measurements of objects in the halo of the Milky Way.  Such observations will be sensitive to a small-scale cutoff in the halo mass function between $10^{6}$ and $10^{8}$~M$_{\odot}$, the density structure within (nearly) pure dark matter halos, and interactions between dark matter particles and either other dark matter particles or Standard Model species.

\textbf{High resolution imaging of strong lenses.}
Constraining dark matter using strong lens systems requires high-sensitivity and high-angular resolution imaging and spectroscopy. For the gravitational imaging technique, comparisons of HST vs. adaptive optics (AO) imaging data from the 10~m Keck telescope have confirmed that sensitivity scales with image resolution. The current AO system on Keck produces a typical angular resolution of 60--90~mas and a Strehl ratio of $\sim$ 10\%--30\% for off-axis targets. With a bigger mirror, a better AO system, and a larger population of potential tip-tilt stars, ELT performance will be significantly better. The AO systems on next-generation telescopes are designed to reach Strehl ratios as high as 90\% in K band. Furthermore, they will include built-in PSF reconstruction software that provides a model of the PSF for data analysis. The combination of improvements in resolution, sensitivity, and PSF quality and knowledge will make the ELTs much more capable than current systems. Specifically, only the brightest handful of lenses in the sky can currently be targeted with 8--10~m class telescopes. The US ELTs, on the other hand, will provide virtually full sky access, allowing us to target large samples of rare quadruply-imaged objects, selecting the ones with the best (i.e., the ones containing the most structure) extended images for the gravitational imaging technique. Moreover, Fisher analyses of simulated data have shown that lens modeling of spectroscopically resolved images can mitigate the source-subhalo degeneracies, dramatically increasing the sensitivity of the observations to low-mass subhalos \citep{Hezaveh:2013}.

For the flux-ratio anomaly technique, a very promising route forward is AO-assisted IFU spectroscopy. Integral field spectroscopy allows the measurement of flux ratios from the narrow-line region of the lensed active galactic nucleus (AGN), which should be immune to microlensing \citep[see][]{Nierenberg:2014, Nierenberg:2017, Nierenberg:2020}. Unfortunately, the technique is currently restricted to only a handful of quad lenses, given the limitations of tip-tilt stars and the need to have the narrow AGN emission lines fall in transparent windows of the Earth’s atmosphere. As in the case of gravitational imaging, a two-hemisphere US ELT system will enable the application of the technique to the much larger samples required to provide definitive conclusions on the nature of dark matter by virtue of their sensitivity (exploiting the $D^{4}$ advantage of large telescopes for observing point sources in diffraction-limited mode), resolution (for astrometry of the images), and full-sky access.

In order to measure the dark matter halo mass function in the range $10^{7}-10^{10}$~M$_{\odot}$, forward-modeling simulations indicate that samples of $\sim$ 50 extended arcs and $\leq$ 50 quadruply-imaged quasars would be needed. The former provide sufficient statistics of the rarer $10^{9}-10^{10}$~M$_{\odot}$ substructures, while the latter are needed for direct and statistical constraints on the mass function in the range $10^{7}-10^{9}$~M$_{\odot}$. For the gravitational imaging technique, the brighter the source galaxy, the more clumpy its structure, and the better the separation between the lens galaxy and the background lensed object, the better the substructure constraints will be. The sample of lenses should contain low and high redshift lenses (and sources) to statistically break the degeneracy between structure along the line of sight and subhalos bound to the main deflector and characterize possible evolutionary trends. Gravitational imaging will also be able to provide constraints on the inner density slope of the massive subhalos ($\sim 10^{10}$~M$_{\odot}$), while the flux ratios of lensed quasars are sensitive to the dark matter self-interaction cross-section in addition to the mass function \citep{Gilman2021}. Observations of each lens system are expected to require integration times of a few hours on an ELT.

\begin{figure}[th]
    \centering
    \hspace*{-0.53cm}\includegraphics[width=1.0\textwidth]{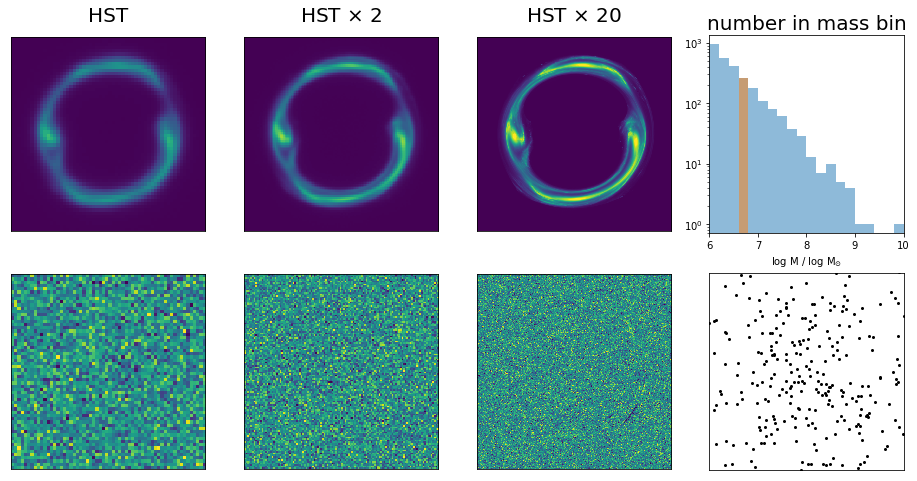}
    \caption{Illustration of the effect of angular resolution on sensitivity of a gravitationally lensed arc to dark matter substructure.  The top row shows images of a galaxy lensed into a near-Einstein ring simulated with the \textsc{lenstronomy} package \citep{lenstronomy}, and the bottom row displays the per-pixel residuals between simulations where the lens includes substructure or does not.  In the left column, the resolution is that provided by the Hubble Space Telescope (HST).  In the middle left column, the resolution is twice that of HST, corresponding to the use of adaptive optics on a current telescope.  In the middle right column, the resolution is increased by another factor of 10, comparable to what is expected for diffraction-limited ELT imaging.  Only at this resolution is the substructure signal detectable in the bottom row.  The far right column shows the full mass function of substructures expected in $\Lambda$CDM in the top panel, where the single mass bin included in the simulation is highlighted, and the spatial distribution of the substructures in the bottom panel.}
    \label{fig:lensing_elts}
\end{figure}

\textbf{Astrometry and spectroscopy of Milky Way satellite galaxies.}  Complementing measurements of the mass function of dark matter subhalos around distant galaxies from lensing, the dwarf galaxies orbiting the Milky Way provide the best probe of the subhalo mass function above $\sim10^{8}$~M$_{\odot}$ in the local universe.  The mass contained within the half-light radius of a dwarf can be accurately determined from the velocity dispersion of the stars in the galaxy \citep{Wolf2010}.  Although the mass contained within this central region is only a small fraction of the total halo mass, the mass function can be recovered from a well-defined sample of high-quality kinematic measurements \citep[e.g.,][]{Nadler2020}.  However, given the large number of new dwarfs expected to be identified by the Rubin Observatory, and the typical distances and luminosities of those systems, it will not be possible to obtain a complete sample of mass measurements with existing spectroscopic resources \citep{Simon2019}.  The increased sensitivity provided by ELTs, especially if combined with relatively wide-field spectroscopic instruments, will be required to observe all of the Rubin dwarf galaxies.

ELTs will also enable the first unambiguous measurements of the dark matter density profiles of the lowest-mass dwarf galaxies.  There is an extensive literature of density profile measurements for larger dwarf galaxies stretching back more than two decades \citep[e.g.,][]{Flores1994,deBlok2001,Oh2011,Relatores2019}, but the dark matter distribution at those mass scales is thought to be altered by feedback processes \citep[e.g.,][]{Governato2012}.  In much lower-mass galaxies such as the Milky Way satellites, stellar feedback should not be strong enough to grossly affect the density profile.  Unfortunately, the most recent modeling indicates that clearly distinguishing central dark matter cusps from cores in Milky Way satellites cannot be achieved with radial velocity measurements alone \citep{Guerra2021}.  As originally suggested by \cite{Strigari2007}, though, combining proper motion measurements of dwarf galaxy stars with radial velocities will break the degeneracies that have plagued previous studies.  The $\sim20$~$\mu$as astrometric precision of ELTs will make it possible to detect the motions of dwarf galaxy stars as they orbit within their host galaxies over a time baseline of five years \citep{Simon2019b}.  ELTs will also provide the sensitivity needed to achieve this precision at faint enough magnitudes to provide large stellar samples.  In concert with existing or new radial velocity data sets, ELT astrometry will clearly distinguish cuspy and cored density profiles, offering a crucial test of dark matter models.

\textbf{Time-series extreme precision radial velocity observations (``optical timing").}  Traditionally, kinematic modeling has been used to estimate Galactic accelerations, and thereby the mass distribution of the Milky Way.  In a time-dependent potential like that of our Galaxy, kinematic modeling can lead to an inaccurate inference of the Galactic mass distribution  \citep{Haines2019}.  Measurements of the  accelerations of stars that live within the gravitational potential of the Galaxy have recently become feasible due to advances in technology, and in principle provide the most precise characterization of the dark matter distribution in the Galaxy.  Because this approach requires extreme-precision time-series measurements, we denote it ``optical timing" in analogy with pulsar timing.  Over decade baselines, the expected Galactic acceleration, i.e., the change in the line-of-sight velocity, $\Delta RV$, experienced by stars at $\sim$ few kpc distances from the Sun is $\sim$ 10 cm~s$^{-1}$~decade$^{-1}$ \citep{SilverwoodEasther2019,Ravi2019,Chakrabarti2020}.  The current generation of stabilized spectrographs such as NEID \citep{Schwab2016} and ESPRESSO \citep{Pepe2021} are approaching this level of RV precision \citep[e.g.,][]{Netto2021}, but because they are on 3.5 m and 8 m telescopes respectively, they only allow for measurements of accelerations of relatively bright stars within a few kpc of the Sun.  

Future instruments, such as G-CLEF \citep{Szent2017} on the GMT and MODHIS on the TMT \citep{Mawet2019}, are expected to allow much greater dynamic range and enable direct acceleration measurements across the Galaxy.  Like pulsar timing, these observations will yield a direct measurement of Galactic accelerations, which provides the mass distribution (both the smooth component and dark matter substructure) without the assumptions of equilibrium or symmetry inherent in traditional kinematic analyses. It should be possible to characterize the upper mass range of the dark matter halo mass function ($10^{9} - 10^{10}$~M$_{\odot}$) with a sample of tens of ``quiet" stars (i.e., stars with low enough RV jitter such that those intrinsic stellar velocity fluctuations do not contaminate the Galactic signal \citep{WrightRobertson2017}).  Moreover, one has to ensure that such sources trace the Galactic signal and are not contaminated by binaries,  planets, or intrinsic stellar variability (``stellar jitter") \citep{Chakrabarti2020}.  Perhaps the most serious contaminating factor to directly measure Galactic accelerations down to the $\sim$ 10 cm~s$^{-1}$~decade$^{-1}$ level with EPRV observations is stellar jitter.  Recent analyses suggest that machine learning techniques may be able to mitigate this to some extent \citep{Debeurs2020}.  Analysis of publicly available detailed RV observations of the Sun may also provide a means of better understanding stellar jitter \citep{Lin2021}.  Observations in the near-infrared should also help to mitigate intrinsic RV variability \citep{Mawet2019}.  The current generation of spectrographs have demonstrated significant gains in precision, and enough so to make a compelling argument to measure Galactic accelerations.  The main technical challenge that now needs to be overcome is the demonstration of stability over decade timescales that can allow for such a measurement. Relative to current extreme-precision radial velocity (EPRV) surveys that can be conducted by ESPRESSO and existing spectrographs, the ELTs will allow for not only greater coverage in volume, but also should allow for a denser sampling of sources (hundreds) and thereby constraints on lower-mass dark matter subhalos, perhaps extending as far as $\sim 10^{6}$~M$_{\odot}$.  A key point here is that EPRV is equally well a tool for dark matter science as it is for detecting and characterizing exoplanets.

\subsection{Wide-field Multi-Object Spectroscopy Facilities}
\label{sec_wfmos}

\begin{figure*}
\includegraphics[width=0.49\textwidth]{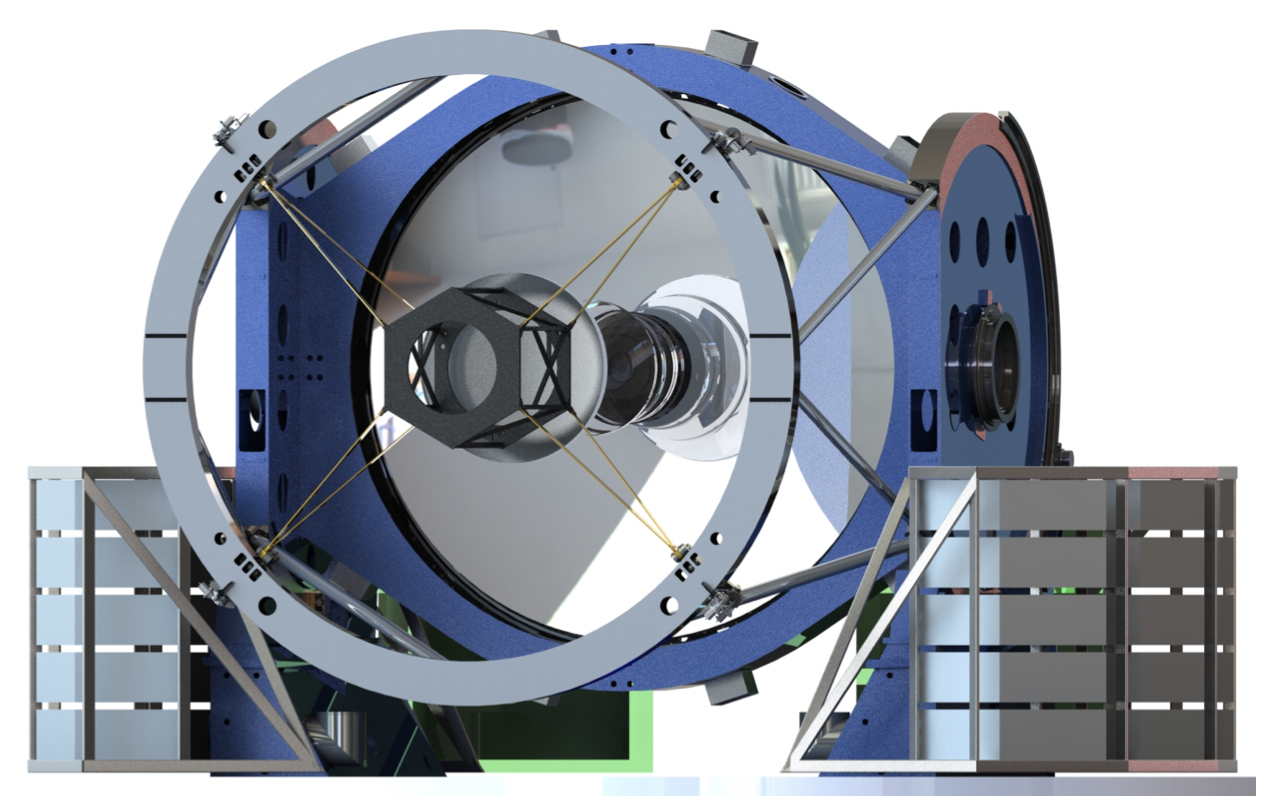}
\includegraphics[width=0.49\textwidth]{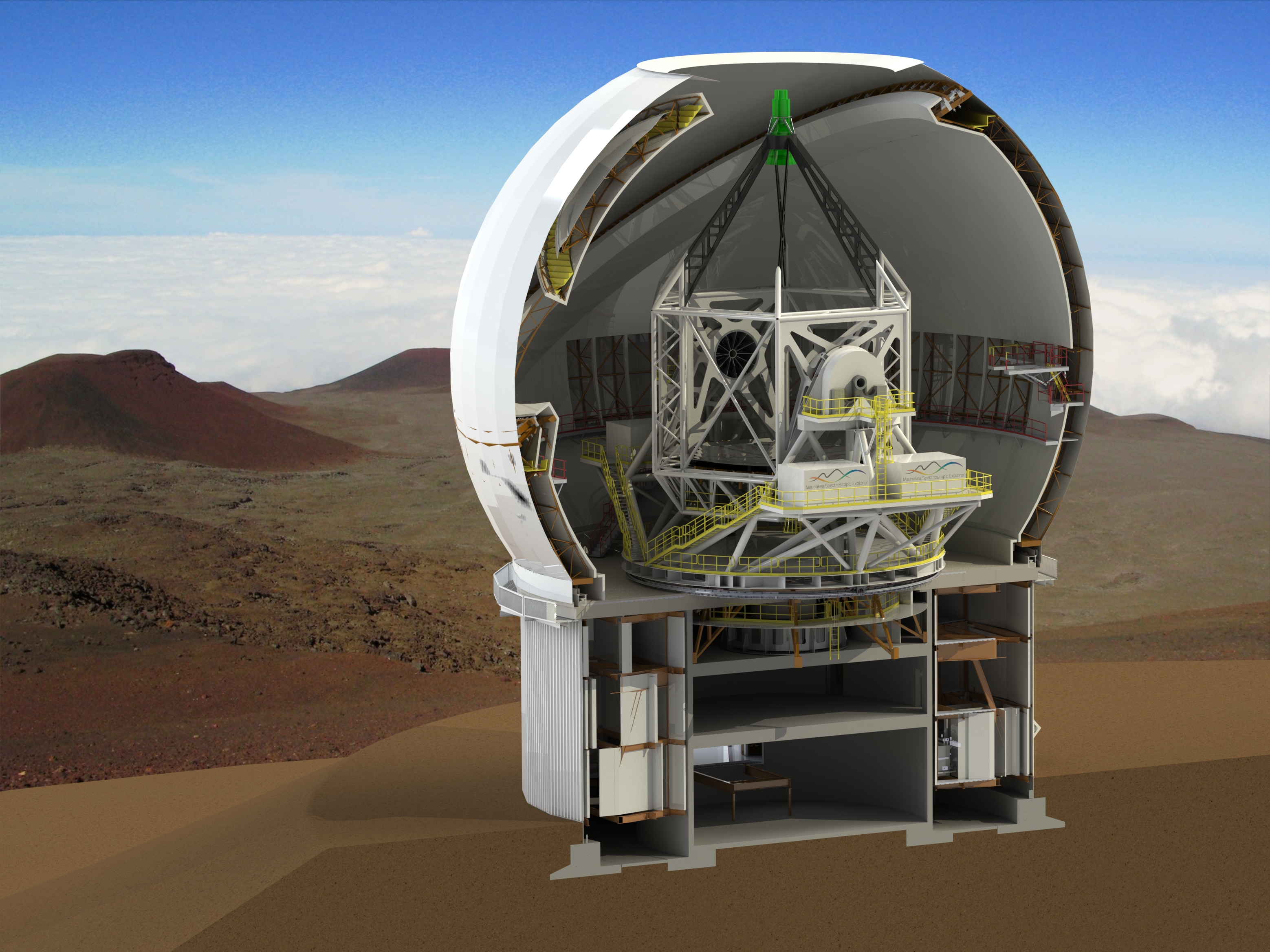}
\caption{\label{fig:wfmos} Conceptual designs of MegaMapper (left; \citep{schlegel19}) and the Maunakea Spectroscopic Explorer (right; \citep{MSE_book2018}). These are two of the proposed wide-field multi-object spectroscopic facilities.}
\end{figure*}

A critical missing link in the international portfolio of astronomical observatories for the next decade is a dedicated wide-field multi-object spectroscopy (WFMOS) facility on an 8-meter class telescope.  As described in more detail in Section~\ref{sec:optical_measurement_opportunities}, highly-multiplexed spectroscopy on a large telescope offers significant potential for constraining the dark matter particle mass and interaction cross-section, as well as the properties of the Milky Way's dark matter halo.  Several WFMOS facilities on 2.5- to 4-meter telescopes are either already operating or expected to begin in the near future, including but not limited to:
\begin{itemize}[itemsep=0pt]
    \item The SDSS-V survey on two 2.5-meter telescopes \citep{Kollmeier2017}
    \item The DESI survey on the 4-meter Mayall telescope \citep{desi2016}
    \item The 4MOST survey on the 4-meter VISTA telescope \citep{deJong2011}
    \item The WEAVE survey on the 4-meter William Herschel Telescope \citep{Dalton2016}.
\end{itemize}

\noindent
However, due to their aperture size, these facilities have limited capacity to observe the faint objects that will be the highest-priority targets identified by the Rubin Observatory.  In contrast, the Prime Focus Spectrograph (PFS) on the 8.2-meter Subaru Telescope will provide similar multiplexing and a larger aperture \citep{Takada2014}, but it does not offer access to most of the US community, it will share time on the telscope with other Subaru instruments, and it has limited sky overlap with the Rubin survey area given its location in Hawaii.  Thus, although the need for such a capability has been broadly agreed upon by the international community (see below), there is currently no dedicated next-generation WFMOS facility that has been funded or approved for construction.

In particular, a US community study supported by NSF and the Kavli Foundation included as one of its top recommendations that the US ``[d]evelop or obtain access to a highly multiplexed, wide-field optical multi-object spectroscopic capability on an 8m-class telescope" \citep{Najita2016}.  The following year, a working group from the European Southern Observatory came to a similar conclusion: ``We consider the scientific case for a large aperture (10-12m class) optical spectroscopic survey telescope with a field of view comparable to that of LSST. We find that such a facility could enable transformational progress in several broad areas of astrophysics, and may constitute an unmatched ESO capability for decades" \citep{Ellis2017}.  Most recently, the Astro2020 Decadal Survey identified highly multiplexed spectroscopy as a strategic priority.  It emphasized the need for new capabilities in highly multiplexed spectroscopy to complement the Rubin Observatory and to advance the scientific priorities of the survey.
Astro2020 recommended that a major (MSRI-2 scale) investment could be made in a large, dedicated facility late in the coming decade.  However, because the NSF MSRI-2 cost cap is $\sim\$100$~million, partnerships with other organizations or agencies are likely to be required to complete such an ambitious project.

Here we describe several proposed (but not funded) next-generation WFMOS facilities that could make the key measurements listed in Section~\ref{sec:optical_measurement_opportunities}.

\subsubsection{DESI-II}
\label{sec:desi2}
DESI (\S~\ref{sec:desi}) is an ambitious multi-fiber optical spectrograph using 5000 robotically-positioned fibers in an 8 deg$^2$ focal plane on the Kitt Peak National Observatory Mayall 4-meter telescope. Fibers feed a bank of 10 triple-arm spectrographs that measure the full bandpass from 360 nm to 980 nm at a spectral resolution of $R \sim 2000$ in the UV and over $R \sim 4000$ in the red and near-infrared.  The DESI five-year survey will probe dark energy and cosmic structure by obtaining redshifts for more than 35 million galaxies and quasars at $z<3.5$ over 14,000 deg$^2$ of the sky \citep{desi2016}.
In addition to the galaxy and quasar survey, DESI will observe roughly ten million stars in the Milky Way.  Because these observations will be conducted during brighter conditions, the stellar targets are selected over the magnitude range $16 < r < 19$, ensuring a sufficient signal-to-noise ratio for measurements of radial velocities and stellar parameters.

Beyond the end of its planned survey, DESI will remain a state-of-the-art facility for wide-field surveys. Given the field of view, multiplex, throughput, and resolution, a second phase of DESI (DESI-II) offers the potential for exciting scientific opportunities for studying dark matter in the 2026-2031 time frame. For example, DESI-II could dedicate a significant amount of dark time to spectroscopic targets for dark matter science. The observation of fainter stars (e.g., pushing from $\sim19$ mag in DESI to $\sim 21$ mag in DESI-II) in the Milky Way will probe the dark matter halo distribution in our Milky Way at both small and large scales \citep{Dey2019BAAS}. Increasing the sample size of the faintest, low redshift galaxies will improve the constraints of the satellite luminosity function \citep{saga2}.

Although DESI will still be a premier spectroscopic facility in its current configuration, there may also be opportunities for augmentation. For example, the spectrographs could be upgraded to include  Ge CCDs for near-infrared wavelength coverage up to 1.3~$\mu$m, or with ultra-low noise Skipper CCDs to improve performance at blue wavelengths \citep{Drlica-Wagner:2020}. 
Not including any major upgrades, the cost to operate a second-generation DESI survey is likely to be $\sim$\$10--15 million per year \citep{Levi2019BAAS}.

\subsubsection{MegaMapper}
\label{sec:megamapper}
MegaMapper is a very wide-field 6.5-m telescope in the initial phases of facility design.  With the exception of modifications to the primary mirror to greatly increase the field of view, the telescope will be a copy of the two Magellan telescopes.  This shared design takes advantage of Magellan's successful engineering heritage and two decades of operational experience.  The current optical design for the system offers a field of view of $\sim7$\,deg$^{2}$.  The envisioned instrument to be paired with the telescope consists of a bank of medium-resolution spectrographs fed by 20,000 optical fibers, although a wide variety of other spectroscopic configurations are also possible.  The initial projection for the construction cost is approximately $\$160$~million.  The telescope construction may be funded by NSF through the MSRI program, as suggested by the Astro2020 decadal survey report, while the spectrographs, detectors, fiber positioners, and front-end electronics present attractive opportunities for potential DOE contributions, building on the success and technical capabilities pioneered by DESI.

With its combination of substantially larger aperture, wider field of view, and significant increase in spectroscopic multiplexing, MegaMapper offers a survey speed exceeding any other existing facility by an order of magnitude \citep{schlegel19}.  As described in several Snowmass Letters of Interest, as well as documentation provided to Astro2020, MegaMapper will conduct a galaxy redshift survey of $\sim10^{8}$ galaxies.
These measurements will provide the tightest available constraints on non-Gaussianity and measure the fraction of dark energy at the $<1\%$ level to $z=4.5$.

Through either simultaneous observations or a separate dedicated program, MegaMapper could also enable important new constraints on dark matter models.  Specifically, precise kinematic measurements for enormous samples of stars in nearby dwarf galaxies and stellar streams will probe the 
mass function of dark matter halos below $10^{8}$~M$_{\odot}$, set new limits on the mass of warm dark matter and fuzzy dark matter particles, and constrain the interaction cross-section of dark matter particles with one another and with Standard Model species (see halo WP).  In addition, velocities of stars in streams and the halo of the Milky Way will significantly improve mass measurements of the Milky Way, which are critical to a broad set of direct and indirect detection experiments, as well as cosmic probes of dark matter.

\subsubsection{Maunakea Spectroscopic Explorer}
\label{sec:mse}
The Maunakea Spectroscopic Explorer (MSE) will transform the CFHT 3.6m optical telescope into an 11.25 meter aperture telescope to feed 4,332 fibers over a wide 1.52~deg$^{2}$ field of view, based on its 2018 Conceptual Design (Figure~\ref{fig:wfmos}; \citep{MSE_book2018}).  In its current design, among the 4332 fibers,  2166 of them will be fed to 3 low resolution spectrographs with a wavelength coverage of 0.36--1.00 $\mu$m and a spectral resolution of $R\sim5000-6000$, 1083 fibers will be fed to 2 infrared spectrographs with a wavelength coverage of 1.00 - 1.80 and a spectral resolution of $R\sim5000-7000$, and the remaining 1083 fibers will be fed to 2 high-resolution spectrographs with a spectral resolution of $R\sim30000$ at smaller wavelength coverage. Furthermore, trade studies are currently underway to investigate significantly increasing the number of fibers (by a factor of $\sim5$) in the focal plane by implementing a quad-mirror, Rubin Observatory-style telescope design, with a 14-meter aperture primary mirror. MSE has completed its Conceptual Design; since that time it has been advancing the instrument design and is currently preparing to begin a two-year Preliminary Design Phase in late 2022, with science operations expected to begin in 2032. 

MSE has assembled a Detailed Science Case document~\cite{mse_dsc}, of which one chapter focuses on the concrete ways that astrophysical probes can elucidate the particle nature of dark matter with MSE \citep{mse_dm}. The dark matter-related science cases range from determining the line-of-sight velocities of a large number of faint stars in the Milky Way's stellar streams and nearby dwarf spheroidal galaxies, obtaining the redshifts of low-mass galaxies in the local Universe ($z<0.05$), to searching for strongly lensed galaxies at higher redshift. N-body and hydrodynamical simulations of cold, warm, fuzzy and self-interacting dark matter show that non-trivial dynamics in the dark sector will leave an imprint on structure formation.
Sensitivity to these imprints will require extensive and unprecedented kinematic datasets for stars down to $r \sim 23$ mag and redshifts for galaxies down to $r \sim 24$ mag. A 10m class wide-field, high-multiplex spectroscopic survey facility like MSE is required in the next decade to provide a definitive search for deviations from the cold collisionless dark matter model. 

MSE is expected to build on the existing Canada-France-Hawaii collaboration with more international partners including Australia, China, India, South Korea, and the US and UK astronomical communities. The total construction cost of MSE is about \$450 million and the expected funding model involves cost sharing between partners, $\sim30\%$ of which is expected from the US.

\subsubsection{SpecTel}
\label{sec:spectel}

SpecTel is a spectroscopic facility proposed for construction in Chile, largely motivated by the ESO working group response to community needs as described in the beginning of Section~\ref{sec_wfmos}. 
The conceptual design of the telescope includes an 11.4-meter aperture with a 4.91 square degree field of view and a focal plane at the Cassegrain focus well-suited for fiber-fed spectroscopy.
The focal plane has a 1.43 meter diameter, enabling instrumentation of 15,000 robotically-controlled fibers using the 10.4 mm pitch design from DESI \citep{SpecTel:2017,SpecTel:2019}.

New fiber positioner technologies such as the 6.5 mm pitch positioners proposed for the MegaMapper facility would allow a significant increase in multiplex number.
The Cosmic Visions panel suggested a goal of a 5-mm pitch for a new fiber positioner technology \cite{cosmicvisions2018}.
Reaching that spacing would allow simultaneous spectroscopy from 60,000 fibers, enough for a program that covers 10\% of the galaxies in the LSST Gold Sample.
Research and Development effort for new positioners has begun at DOE labs and could be a major contribution from the US community toward instrumenting SpecTel or any other spectroscopic facility.
The conceptual design for SpecTel provides an optical quality that would allows spectral wavelength coverage over a range of $360< \lambda < 1330$ nm.
Ge CCD development for the IR portion of this range and expertise in spectrograph design and construction could also be be significant US contributions to such a facility.
Finally, the expertise from DESI in spectroscopic pipeline development and data management could be directed toward a future spectroscopic program as a US contribution.

The science cases for SpecTel identified by the ESO working group are very similar to those in MSE. The consistency in observational programs, including those that would provide new insights into dark matter, further highlight the need for a wide-field, highly multiplexed spectroscopic facility. The primary design difference between SpecTel and MSE lies in the optical path. The Cassegrain focal plane in the preliminary SpecTel design includes only a three-lens transmissive element that serves as both field flattener and atmospheric differential corrector. The most recent MSE design relies on only reflective surfaces to minimize achromatic effects and increase the potential wavelength coverage.

The ESO community is pursuing a conceptual design study for SpecTel which would be required to establish the cost and the construction schedule.  Preliminary cost estimates (FY2019) for the telescope and enclosure were \$200M. Additional costs for the instrument would depend on the scope of the spectroscopic system, where both medium resolution and high resolution options are being considered.

\subsubsection{Comparison and Summary}

\begin{table*}
\small
\centering
\begin{tabular}{|p{6cm}|p{2.0cm}|p{2.5cm}|p{2.2cm}|p{2.2cm}|}
\hline
    Wide-field MOS Facilities/Surveys &  DESI-II & MegaMapper & MSE & SpecTel\\
    \hline
    Telescope aperture diameter (m) &  4 &  6.5 &  11.25--14 & 11.4 \\
    Field-of-view (deg$^{2}$)             &  8.94 &  7.06 &   1.52 & 4.91 \\
    Number of fibers                    &  5000 & 20000  & 4332--21000 & 15000--60000\\
    Spectral resolution $\lambda/\delta\lambda$  &  3000 & 4000 & 5000/20000 & 4000/40000 \\
    Wavelength range ($\mu$m)  &  0.36-1 & 0.36-1 & 0.36-1.8 & 0.36-1.3 \\
    Timescale/first light & 2026 & 2030 & 2032 & $>$2035 \\
    Total project cost                                &  low  & medium  & high & high\\
    Expected US contribution (in \$ million)                     &  $<50$  & 160  & 135 & ???\\
    Site location                        & US/Arizona  & Chile  &  US/Hawai'i  & Chile \\
    \hline
\end{tabular}
\caption{Summary and Comparison of the Key Parameters of Four WFMOS Facilities Discussed in this White Paper.  }\label{mos_compare}
\end{table*}

The scientific need for a next-generation massively-multiplexed spectroscopic facility on a large telescope is clear.  Such a project will not only make a major impact on dark matter physics, but the same facility would also enable tremendous progress on studies of dark energy, inflation, neutrino masses, and a broad set of astrophysics.  Above we have described four candidate facilities that could achieve these goals in different forms, but this list is not intended to be exclusive; for example, the FOBOS spectrograph for Keck \citep{Bundy2020} is a smaller scale instrument that could contribute to some of the same science. In addition, an ELT spectrograph offering higher multiplexing than their currently-planned instruments could be competitive for some of the WFMOS science goals because of the large collecting area of these telescopes.

In Table~\ref{mos_compare}, we summarize the key parameters of the WFMOS facilities discussed in this section, and below we elaborate on this comparison.  

The scientific capabilities of these facilities are a function of four main parameters: the telescope collecting area, its field of view, the number of fibers, and the spectrograph characteristics.  According to the currently available designs, the greatest sensitivity is offered by MSE and SpecTel thanks to their larger primary mirrors, while DESI-II and MegaMapper provide the widest fields of view. However, all of the projects are in early stages of design so that significant design changes are still possible.
For example, a range of options is presented for fiber number for MSE and SpecTel, and MSE may consider a larger telescope diameter as well.  We note that the importance of field of view relative to multiplexing depends on the surface density of the targets of interest.  Choices about the number of fibers for each facility were generally driven by science goals unrelated to dark matter, and the multiplexing of these facilities should not be the limiting factor for any of the measurement opportunities discussed in Section~\ref{sec:optical_measurement_opportunities}.

Any one of the three new telescope projects can accomplish the dark matter science goals described above.  There are differences in detail among their capabilities, which will enable certain measurements to be made more efficiently, or dark matter limits to be extended further, with some facilities.  For example, the higher spectral resolution offered by MSE will enable more accurate stellar velocity measurements, potentially constraining the dark matter subhalo mass function down to $10^{5}$~M$_{\odot}$, compared to the $10^{6}$--$10^{7}$~M$_{\odot}$ limits obtainable at lower spectral resolution \citep{mse_dsc}.  
For similar reasons, a high resolution option is being considered for SpecTel. The larger field of view of MegaMapper would facilitate mapping of stellar streams more quickly, while the larger apertures of MSE and SpecTel can provide larger samples of faint stars in both streams and dwarfs.  However, quantitative estimates for the impact of depth and field of view on dark matter constraints are not yet available.

The locations of the various WFMOS facilities will also play a role in their ability to constrain dark matter parameters.  MegaMapper and SpecTel will have complete overlap with the Rubin Observatory LSST footprint, so that they are able to follow up all dwarf galaxies and streams identified in the LSST data.  On the other hand, MSE will only be able to observe $\sim50$\%\ of the Rubin survey area, but can see a larger fraction of the sky, including northern targets that are inaccessible from Chile.

Finally, there are large differences between projects in cost, and perhaps time scale.  At one extreme, DESI is already operating, and therefore only continuing operation funds are needed for DESI-II.  DESI-II is expected to start operating in 2026 after the conclusion of the current 5-yr DESI survey (2021-2026).  By relying on established hardware and technology as much as possible, MegaMapper aims to minimize the project cost and technical risk, and will be the least expensive of the three proposed facilities.  MSE and SpecTel are projected to be factors of $\sim3$--5 more expensive, although because they are based on international partnerships that are still being developed, the US contribution remains uncertain.

The bottom line is that construction of one of the massively multiplexed spectroscopic facilities will be important for the next generations of cosmic constraints on dark matter.  From the perspective of dark matter science alone, the choice of which facility the US should pursue is less critical.  Cost, schedule, risk, and the needs of other scientific areas in the Cosmic Frontier may reasonably drive that decision.

\section{Future Gravitational Wave Facilities}
\label{sec:gw}

\subsection{Key Measurement Opportunities}

\begin{figure}[th]
    \centering
    \hspace*{-0.53cm}\includegraphics[width=0.68\textwidth]{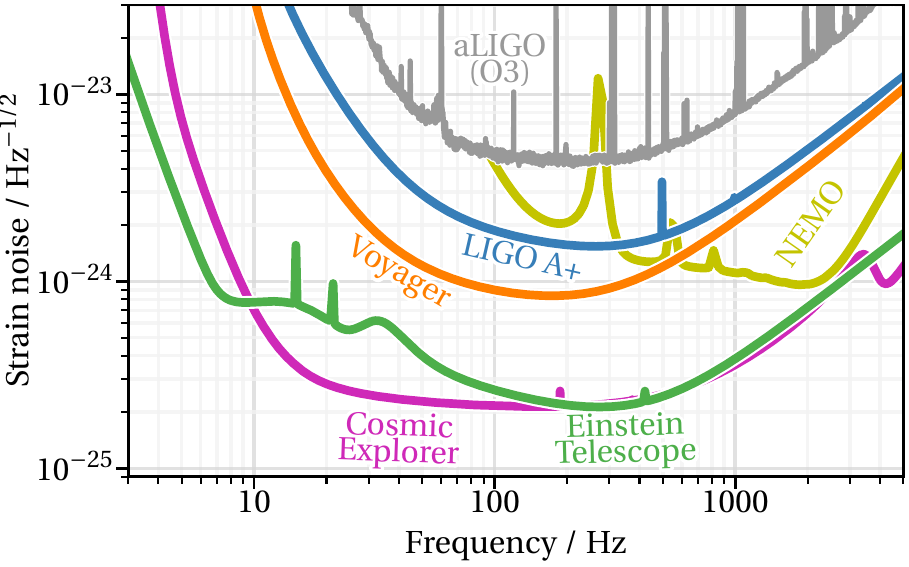}
    \caption{Amplitude spectral densities of detector noise for Cosmic Explorer, the current (O3) and upgraded (A+) sensitivities of Advanced LIGO, LIGO Voyager, the proposed Australian NEMO detector, and the three paired detectors of the triangular Einstein Telescope. At each frequency the noise is referred to the strain produced by a source with optimal orientation and polarization.}
    \label{fig:noise_curves}
\end{figure}

The interaction of dark matter with astrophysical compact objects (black holes and neutron stars) allows gravitational wave observatories to constrain a wide variety of dark matter candidates including axion-like particles, ultralight bosons, and primordial black holes. Cosmic Explorer, a next-generation gravitational-wave observatory that is planned construction in the United States, will yield a factor of ten improvement over the sensitivity of Advanced LIGO~\cite{Reitze2019,Evans:2021gyd}. A possible upgrade to the LIGO facilities, known as LIGO Voyager~\cite{LIGO:2020xsf}, is also under considerations. Figure~\ref{fig:noise_curves} shows the sensitivity of Cosmic Explorer and Voyager compared to the current Advanced LIGO detectors (at their O3 sensitivity), the funded LIGO A+ upgrade, and the Einstein Telescope, a planned European next-generation detector~\cite{ET}. Some of the key measurements that next-generation detectors could provide include:
\begin{enumerate}
    \item \textbf{Searches for axionlike particles.} Light axions inside neutron stars can modify the inspiral waveform yielding a dark matter signature detectable in the gravitational waveform. Constraints can be placed on axions with masses below $10^{-11}$eV and decay constants ranging from $10^{16}$~GeV to $10^{18}$~GeV, with next-generation facilities improving current constraints by a factor of $\sim 3$~\cite{Zhang:2021mks}.
    \item \textbf{Superradiant instabilities of rotating black holes.} Spinning black holes can superradiantly amplify excitations in a surrounding field, generation long-lived ``bosonic clouds'' that slowly dissipate energy through the emission of gravitational waves~\cite{Arvanitaki:2014wva,Arvanitaki:2016qwi}. A large number of unresolved sources can contribute to the stochastic gravitational wave background, allowing Cosmic Explorer to constrain bosons in the mass range $\sim [7 \times 10^{-14}, 2 \times 10^{-11}]$~eV~\cite{Yuan:2021ebu}. In addition to direct detection, the boson cloud spins down the black hole to a characteristic spin determined by the boson mass and the black hole mass. With the large number of high signal-to-noise ratio events that will be seen by Cosmic Explorer, the existance of noninteracting bosons in the mass range $10^{-13}$ to $10^{-12}$~eV could be confirmed through their imprint on the black hole spin distribution~\cite{Ng:2019jsx}.
    \item \textbf{Boson Stars.} Next-generation detectors could distinguish between binaries containing boson stars from binary neutron stars and binary black holes by measuring the compactness parameter of the stars, which typically lies between that of black holes and neutron stars~\cite{Sennett:2017etc,Toubiana:2020lzd}.
    \item \textbf{Primordial Black Holes.} Primordial black holes (PBHs) have been of longstanding candidate for some fraction of dark matter. Binaries containing containing sub-solar mass black holes or black hole binaries observed at very large red shifts would indicate the presence of PBHs as a dark matter component~\cite{Nakamura_1997,Bird:2016dcv}. Primordial black holes could be surrounded by spikes of particle dark matter which could influence the dynamics of a binary merger allowing Cosmic Explorer to probe dark matter candidates with masses heavier than approximately $m_a \sim 10^{-6}$~eV~\cite{Bertone:2019irm}.
\end{enumerate}

\subsection{Cosmic Explorer and Einstein Telescope}
\label{sec:cosmic_explorer}

In the United States, the next-generation Cosmic Explorer facility will have ten times the sensitivity of
Advanced LIGO and will push the reach of gravitational-wave astronomy toward the edge of the observable universe ($z \sim 100$)~\cite{Reitze2019,Evans:2021gyd}. A complementary observatory, known as Einstein Telescope, is under development in Europe. Cosmic Explorer's order-of-magnitude sensitivity improvement will be realized using a dual-recycled Fabry--P\'{e}rot Michelson interferometer, as in Advanced LIGO.
Cosmic Explorer's increased sensitivity comes primarily from scaling up the detector's length from 4 to 40~km. The longer arms increase the amplitude of the observed signals with effectively no increase in the detector noise.   The Cosmic Explorer Horizon Study has considered a reference concept of a 40~km detector and a 20~km detector, both located in the United States. The sensitivity of Cosmic Explorer provides access to significantly more sources, spread out over cosmic time, as well as high-fidelity measurements of strong, nearby sources. These precision measurements will enable the exploration of dark matter candidates through their interaction with compact objects and gravity. The Einstein Telescope is proposed as a 10~km triangular interferometer with $60^\circ$ angles, to be build underground to minimize low-frequency noise. Both Cosmic  the target is data taking in the mid 2030s.

\subsection{LIGO Voyager}
\label{sec:ligo_voyager}

Beyond the A+ upgrade, LIGO is exploring an future upgrade that uses cryogenic silicon optics and suspensions and reduces quantum and Newtonian (gravity gradient) noise; this upgrade is known as Voyager~\cite{LIGO:2020xsf}. LIGO-Voyager represents the best path to maximizing the sensitivity of the existing LIGO facilities, although it does not reach the sensitivity possible in a new facility like Cosmic Explorer or Einstein Telescope. The Voyager technologies are complementary to those being explored for next-generation detectors. Options for lasers, photodiodes, electro-optics for Voyager's planned 2 $\mu$m operating wavelength, and cryogenic engineering solutions to cool the suspended optics will be tested in the Caltech 40-m prototype interferometer and by the European Einstein Telescope pathfinder prototype.

\section{Conclusion}

The design and requirements of observational facilities for studying dark matter overlap heavily with facilities that measure dark energy and inflation. Furthermore, these facilities leverage the core capabilities and expertise of the HEP community.
For example, the HEP community has demonstrated extensive capability in massively multiplexed spectroscopy through SDSS, BOSS, eBOSS, and DESI. Furthermore, DOE has been extensive involved in detector development and packaging for DES, DESI, and Rubin, as well as R\&D on future ultra-low noise Skipper CCDs for future spectroscopic surveys. Current efforts in detector fabrication and readout systems that are being developed for CMB-S4 will translate directly to future CMB experiments.
Additional efforts are already underway to develop radio capabilities. The development of next-generation gravitational-wave observatories will open new windows on dark matter both directly and through their interactions with astrophysical compact objects. The HEP national laboratories provide extensive expertise in cryogenics, mechanical engineering, readout electronics, precision metrology, vacuum technology, project management, and large-scale computing.

Observational facilities are a necessary avenue to explore the fundamental nature of dark matter. For some dark matter models, they will provide important measurements to complement terrestrial experiments, while other dark matter models can only be accessed by observational facilities. The technical and scientific capabilities of the HEP community provide an essential resource for constructing, operating, and analyzing these experiments in the coming decade.

\acknowledgments{We thank Katelin Schutz for providing helpful feedback on a draft of this manuscript.  We are grateful to the CF3 conveners, Alex Drlica-Wagner, Chanda Prescod-Weinstein, and Hai-Bo Yu, for their guidance and management of the Snowmass process.

This research has made use of NASA's Astrophysics Data System Bibliographic Services.}

\bibliographystyle{JHEP.bst}
\bibliography{main.bib}

\end{document}